\documentclass[twocolumn]{aastex63}
\usepackage{natbib}
\usepackage{amsmath} 

\newcommand{\msun}{$M_{\odot}$}
\newcommand{\sersic}{S\'ersic}

\newcommand{\mbh}{$M_{\rm BH}$}

\newcommand{\catnr}{50,000}
\newcommand{\bronzenr}{40}
\newcommand{\silvernr}{26}
\newcommand{\goldnr}{17}

\usepackage[encapsulated]{CJK}

\begin{document}

\title{UNCOVER: Candidate Red Active Galactic Nuclei at $3<z<7$ with JWST and ALMA}
\author[0000-0002-2057-5376]{Ivo Labbe}
\affiliation{Centre for Astrophysics and Supercomputing, Swinburne University of Technology, Melbourne, VIC 3122, Australia}

\author[0000-0002-5612-3427]{Jenny E. Greene}
\affiliation{Department of Astrophysical Sciences, Princeton University, 4 Ivy Lane, Princeton, NJ 08544}

\author[0000-0001-5063-8254]{Rachel Bezanson}
\affiliation{Department of Physics and Astronomy and PITT PACC, University of Pittsburgh, Pittsburgh, PA 15260, USA}

\author[0000-0001-7201-5066]{Seiji Fujimoto}\altaffiliation{Hubble Fellow}
\affiliation{
Department of Astronomy, The University of Texas at Austin, Austin, TX 78712, USA
}
\author[0000-0001-6278-032X]{Lukas J. Furtak}
\affiliation{Physics Department, Ben-Gurion University of the Negev, P.O. Box 653, Beer-Sheva 8410501, Israel}
\author[0000-0003-4700-663X]{Andy D. Goulding}
\affiliation{Department of Astrophysical Sciences, Princeton University, 4 Ivy Lane, Princeton, NJ 08544}
\author[0000-0003-2871-127X]{Jorryt Matthee}
\affiliation{Department of Physics, ET+
H Zurich, Wolfgang-Pauli-Strasse 27, 8093 Zurich, Switzerland}

\author[0000-0003-3997-5705]{Rohan P. Naidu}
\thanks{NASA Hubble Fellow}
\affiliation{Center for Astrophysics $|$ Harvard \& Smithsonian, 60 Garden Street, Cambridge, MA 02138, USA}
\affiliation{MIT Kavli Institute for Astrophysics and Space Research, 77 Massachusetts Ave., Cambridge, MA 02139, USA}

\author[0000-0001-5851-6649]{Pascal A. Oesch}
\affiliation{Department of Astronomy, University of Geneva, Chemin Pegasi 51, 1290 Versoix, Switzerland}
\affiliation{Cosmic Dawn Center (DAWN), Niels Bohr Institute, University of Copenhagen, Jagtvej 128, K{\o}benhavn N, DK-2200, Denmark}

\author[0000-0002-7570-0824]{Hakim Atek}
\affiliation{Institut d'Astrophysique de Paris, CNRS, Sorbonne Universit\'e, 98bis Boulevard Arago, 75014, Paris, France}

\author[0000-0003-2680-005X]{Gabriel Brammer}
\affiliation{Cosmic Dawn Center (DAWN), Niels Bohr Institute, University of Copenhagen, Jagtvej 128, K{\o}benhavn N, DK-2200, Denmark}

\author{Iryna Chemerynska}
\affiliation{Institut d'Astrophysique de Paris, CNRS, Sorbonne Universit\'e, 98bis Boulevard Arago, 75014, Paris, France}

\author[0000-0001-7410-7669]{Dan Coe}
\affiliation{Space Telescope Science Institute (STScI), 3700 San Martin Drive, Baltimore, MD 21218, USA}
\affiliation{Association of Universities for Research in Astronomy (AURA), Inc. for the European Space Agency (ESA)}
\affiliation{Center for Astrophysical Sciences, Department of Physics and Astronomy, The Johns Hopkins University, 3400 N Charles St. Baltimore, MD 21218, USA}

\author[0000-0002-7031-2865]{Sam E. Cutler}
\affiliation{Department of Astronomy, University of Massachusetts, Amherst, MA 01003, USA}

\author[0000-0001-8460-1564]{Pratika Dayal}
\affiliation{Kapteyn Astronomical Institute, University of Groningen, P.O. Box 800, 9700 AV Groningen, The Netherlands}

\author[0000-0002-1109-1919]{Robert Feldmann}
\affiliation{Institute for Computational Science, University of Zurich, Winterhurerstrasse 190, CH-8006 Zurich, Switzerland}

\author[0000-0002-8871-3026]{Marijn Franx}
\affiliation{Leiden Observatory, Leiden University, P.O.Box 9513, NL-2300 AA Leiden, The Netherlands}

\author[0000-0002-3254-9044]{Karl Glazebrook}
\affiliation{Centre for Astrophysics and Supercomputing, Swinburne University of Technology, Melbourne, VIC 3122, Australia}

\author[0000-0001-6755-1315]{Joel Leja}
\affiliation{Department of Astronomy \& Astrophysics, The Pennsylvania State University, University Park, PA 16802, USA}
\affiliation{Institute for Computational \& Data Sciences, The Pennsylvania State University, University Park, PA 16802, USA}
\affiliation{Institute for Gravitation and the Cosmos, The Pennsylvania State University, University Park, PA 16802, USA}

\author[0000-0003-0695-4414]{Michael Maseda}
\affiliation{Department of Astronomy, University of Wisconsin-Madison, 475 N. Charter St., Madison, WI 53706 USA}

\author[0000-0001-9002-3502]{Danilo Marchesini}
\affiliation{Department of Physics and Astronomy, Tufts University, 574 Boston Ave., Medford, MA 02155, USA}

\author[0000-0003-2804-0648]{Themiya Nanayakkara}
\affiliation{Centre for Astrophysics and Supercomputing, Swinburne University of Technology, Melbourne, VIC 3122, Australia}

\author[0000-0002-7524-374X]{Erica J. Nelson}
\affiliation{Department for Astrophysical and Planetary Science, University of Colorado, Boulder, CO 80309, USA}

\author[0000-0002-9651-5716]{Richard Pan}
\affiliation{Department of Physics and Astronomy, Tufts University, 574 Boston Ave., Medford, MA 02155, USA}

\author[0000-0001-7503-8482]{Casey Papovich}
\affiliation{Department of Physics and Astronomy, Texas A\&M University, College Station, TX, 77843-4242 USA}
\affiliation{George P. and Cynthia Woods Mitchell Institute for Fundamental Physics and Astronomy, Texas A\&M University, College Station, TX, 77843-4242 USA}

\author[0000-0002-0108-4176]{Sedona H. Price}
\affiliation{Department of Physics and Astronomy and PITT PACC, University of Pittsburgh, Pittsburgh, PA 15260, USA}

\author[0000-0002-1714-1905]{Katherine A. Suess}
\affiliation{Department of Astronomy and Astrophysics, University of California, Santa Cruz, 1156 High Street, Santa Cruz, CA 95064, USA}
\affiliation{Kavli Institute for Particle Astrophysics and Cosmology and Department of Physics, Stanford University, Stanford, CA 94305, USA}

\author[0000-0001-9269-5046]{Bingjie Wang (\begin{CJK*}{UTF8}{gbsn}王冰洁\ignorespacesafterend\end{CJK*})}
\affiliation{Department of Astronomy \& Astrophysics, The Pennsylvania State University, University Park, PA 16802, USA}
\affiliation{Institute for Computational \& Data Sciences, The Pennsylvania State University, University Park, PA 16802, USA}
\affiliation{Institute for Gravitation and the Cosmos, The Pennsylvania State University, University Park, PA 16802, USA}
\author[0000-0003-1614-196X]{John R. Weaver}
\affiliation{Department of Astronomy, University of Massachusetts, Amherst, MA 01003, USA}

\author[0000-0001-7160-3632]{Katherine E. Whitaker}
\affiliation{Department of Astronomy, University of Massachusetts, Amherst, MA 01003, USA}
\affiliation{Cosmic Dawn Center (DAWN), Niels Bohr Institute, University of Copenhagen, Jagtvej 128, K{\o}benhavn N, DK-2200, Denmark} 

\author[0000-0003-2919-7495]{Christina C. Williams}
\affiliation{NSF’s National Optical-Infrared Astronomy Research Laboratory, 950 N. Cherry Avenue, Tucson, AZ 85719, USA}
\affiliation{Steward Observatory, University of Arizona, 933 North Cherry Avenue, Tucson, AZ 85721, USA}

\author[0000-0002-0350-4488]{Adi Zitrin}
\affiliation{Physics Department, Ben-Gurion University of the Negev, P.O. Box 653, Beer-Sheva 8410501, Israel}

\date{June 2023}

\begin{abstract}
The James Webb Space Telescope (JWST) is revolutionizing our knowledge of $z>5$ galaxies and their actively accreting black holes. Using the \emph{JWST} Cycle 1 Treasury program Ultradeep NIRSpec and NIRCam ObserVations before the Epoch of Reionization (UNCOVER) in the lensing field Abell 2744, we report the identification of a sample of little red dots at $3 < z_{\rm{phot}} < 7$ that likely contain highly-reddened accreting supermassive black holes. Using a NIRCam-only selection to F444W$<27.7$~mag, we find \silvernr\ sources over the $\sim45$ arcmin$^{2}$ field that are blue in F115W$-$F200W$\sim0$ (or $\beta_{\rm UV}\sim-2.0$ for $f_{\lambda} \propto \lambda^\beta$), red in F200W$-$F444W = $1-4$\ ($\beta_{\rm opt} \sim +2.0$), and are dominated by a point-source like central component. Of the 20 sources with deep ALMA 1.2-mm coverage, none are detected individually or in a stack. For the majority of the sample, SED fits to the JWST+ALMA observations prefer models with hot dust rather than obscured star-formation to reproduce the red NIRCam colors and ALMA 1.2~mm non-detections. While compact dusty star formation can not be ruled out, the combination of extremely small sizes ($\langle r_e \rangle\approx50$~pc after correction for magnification), red rest-frame optical slopes, and hot dust can be explained by reddened broad-line active galactic nuclei (AGNs). Our targets have faint $M_{\rm 1450} \approx -14\ \, {\rm to} \, -18$~mag but inferred bolometric luminosities of $L_{\rm bol} = 10^{43}-10^{46}$~erg/s, reflecting their obscured nature. If the candidates are confirmed as AGNs with upcoming UNCOVER spectroscopy, then we have found an abundant population of reddened luminous AGN that are at least ten times more numerous than UV-luminous AGN at the same intrinsic bolometric luminosity.
\end{abstract}


\section{Introduction}

Accreting supermassive black holes (BHs) can be extremely luminous beacons of galaxy formation in the early universe, and over the past twenty years we have been able to detect rare and massive ($\sim 10^9$~\msun) accreting BHs up to $z \sim 7.5$ \citep[i.e. UV-luminous quasars][]{Mortlock:2011,Banados:2018,Fan:2019,Wang:2021,Harikane:2022,Fan:2023}. These rare objects require very wide-area red imaging to find, and have number densities many orders of magnitude lower than typical supermassive BH populations today. Making such monsters from scratch at such early times is a long-standing challenge that has intrigued astronomers for decades \citep[see reviews by ][]{Inayoshi:2020,Greene:2020}. If we could constrain the luminosity function of high-redshift accreting BHs down to low luminosity, that would provide crucial new constraints on when supermassive BHs form and grow \citep[e.g.,][]{ricartenatarajan2018,haimanetal2019,dayal2019,Volonteri:2021,Natarajan:2021,Matsuoka:2022,Haidar:2022,Zhang:2023}. 

\begin{figure*}
\hspace{1.5cm}
\includegraphics[width=0.8\textwidth]{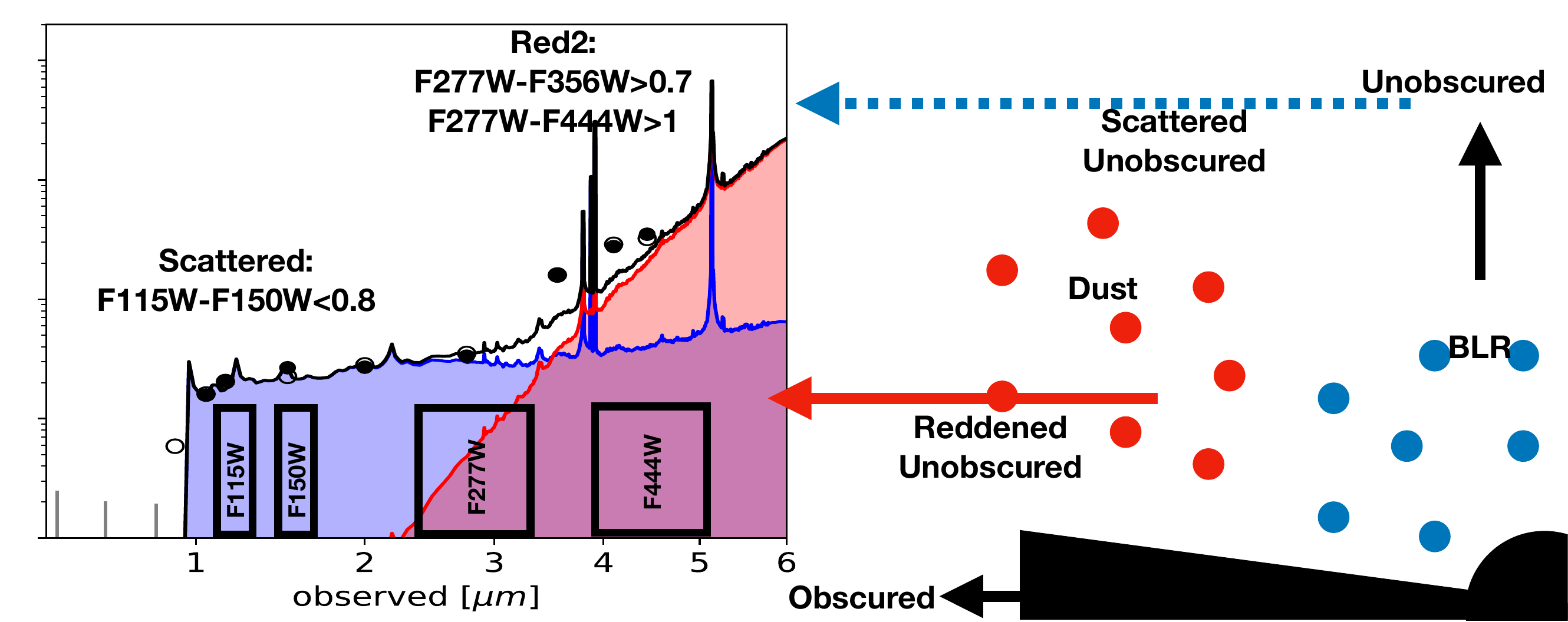}
\caption{Schematic model (right) and photometric fit (left) for compact red sources like CEERS 3210. The model consists of a reddened, but not completely obscured, broad-line AGN template ($red$) representing a direct view of the accretion disk and Broad Line Region, a small contribution from the same AGN template without reddening ($blue$) that we posit represents scattered light, and (3) an unreddened Narrow Line Region with emission line strengths coupled to the broad $H\alpha$ of the red component following typical correlations for broad-line AGN \citep{Stern:2012a,Stern:2012b}. This fit not only provides a decent fit to the photometry, but predicts both broad and narrow line strengths very similar to those observed with NIRSpec.}
\label{fig:ceers3210-phot}
\end{figure*}

The faint-end slope of the quasar luminosity function at $z \gtrsim 6$ has important implications for the cosmic sources of reionization \citep[e.g.,][]{Glikman:2011,Madau:2015,Matsuoka:2018,Giallongo:2019,dayal2020}, for the number density of low-mass black holes \citep[e.g.,][]{Matsuoka:2018,Shen:2019,Onoue:2019}, and for scaling relations between black holes and their host galaxies \citep[e.g.,][]{piana2021, Ding:2022}. Deep optical surveys like the Hyper Suprime-Cam Survey have made progress quantifying the luminosity function down to luminosities of  $M_{1450} \approx -22$ mag or $L_{\rm Bol} \approx 10^{46}$~erg/s at $z>5$ \citep{Matsuoka:2018,Akiyama:2018,Matsuoka:2022}. X-ray surveys reach a bit further, to bolometric luminosities of $L_{\rm bol} \sim 10^{45}$~erg/s at $z \approx 6$ \citep[e.g.,][]{Nanni:2017,Vito:2019,Shen:2020,Wang:2021,Wolf:2022}. Thus current limits reach the Eddington limit for \mbh$\sim 10^8$~\msun, but are not deep enough to probe expected AGN from most drop-out galaxies at these redshifts \citep{VolonteriReines:2016}.

In addition to selection via blue color \citep[e.g.,][]{Richards:2002}, AGN can be selected in many complementary ways. X-rays are less sensitive to obscuration than the UV, and have provided the most complete census of AGN to date at all redshifts \citep[e.g.,][]{Ueda:2014,Aird:2018,Vito:2019,Giallongo:2019}. Mid-infrared selection, likewise, should select sources independent of viewing angle, and has been extremely fruitful in recent years thanks to the combination of \emph{Spitzer} and \emph{WISE} \citep[e.g.,][]{Stern:2005,Goulding:2014,Assef:2018,Zou:2022,Ishikawa:2023}. At the very luminous and rare end are the Hot Dust Obscured Galaxies \citep[HotDOGS][]{Dey:2008,Tsai:2015,Wu:2018} and Extremely Red Quasars \citep[ERQs;][]{Ross:2015,Zakamska:2016,Hamann:2017}. Finally, there are a class of reddened broad-line objects that are challenging to select photometrically \citep[e.g.,][]{Croom:2001}, but have been identified as an important sub-population particularly at the luminous end \citep[e.g.,][]{Glikman:2012,Banerji:2015}. For a nice review, see \citet{Hickox:2018}.

In the era of \emph{JWST}+NIRCAM, we have an amazing new resource for selecting active galaxies. Because of the combination of resolution and sensitivity in the near-infrared bands, we can select high redshift compact sources that are UV-faint but emerging in the rest-frame optical as possible AGN candidates \citep[although there were hints from \emph{HST} and ALMA:][]{Morishita:2020,Fujimoto:2022,Endsley:2023}. NIRCam imaging has already revealed a number of intriguing photometric prospects for low luminosity accreting BHs \citep[e.g.,][]{Onoue:2023, Ono:2022, Furtak:2022, Endsley:2022, Larson:2023}. 

Stunning spectroscopic confirmation of the AGN nature of these sources is now also becoming available \citep{Kocevski:2023,Harikane:2023,Oesch:2023,Ubler:2023,Barro:2023}. \citet{Onoue:2023} identified a point-like source as a likely $z=5.2$ AGN from early Cosmic Evolution Early Release Science (CEERS) Survey data, subsequently found to be a \mbh$\sim 10^7$~\msun\ black hole radiating at $\sim 10\%$ of its Eddington luminosity at $z>5$ from NIRSpec follow-up \citep{Kocevski:2023}. Even more intriguing is the source CEERS 3210. Like CEERS 1670, it has a very compact morphology. However, the spectral energy distribution (SED) changes from blue in NIRCam short-wavelength (SW), with F115W-F200W$\sim0$, to red in NIRCam long-wavelength (LW), with F277W-F444W$\sim2.0$. The object is so red that the rise resembles a Balmer break at $z>7$ \citep{Labbe:2023}. NIRSpec observations also revealed the source to be a broad-line AGN at $z=5.6$ \citep{Kocevski:2023}. \citet{Harikane:2023} and \citet{Barro:2023} present an even larger sample of compact red sources in CEERS, although not all show concrete evidence for broad H$\alpha$ and the nature of these is not always clear. Finally, Matthee et al.\ (2023, submitted) have identified a sample of ``Little Red Dots'' with broad H$\alpha$ in \emph{JWST} grism data \citep{Oesch:2023}.

The central goal of this paper is to photometrically identify more compact red sources with \emph{JWST}/NIRCam imaging, and then use spectral synthesis models to examine the source of the red continuum and UV excess, be it dusty star formation or reddened light from an AGN \ref{fig:ceers3210-phot}. We will propose that the red rest-frame optical continuum likely arises from a reddened broad-line AGN, while the flat blue continuum comes either from an additional scattered component, or some star formation in the host galaxy. Objects like this have been seen very locally \citep[e.g., Mrk~231,][]{Veilleux:2016}, and at a range of redshifts $z<3$ \citep[e.g.,][]{Assef:2020,Pan:2021,Noboriguchi:2022,Glikman:2023}, but it is hard to construct complete samples of reddened broad-line AGN \citep[e.g.,][]{Glikman:2012}. We should note that there are many possible origins for the UV light, including star-formation in the host galaxy or emission from an outflow \citep{Veilleux:2016}, but what matters most for our discussion is the origin of the red rest-frame optical continuum. 

If we could systematically identify this low-luminosity population from deep NIRCam imaging, it would provide critical new insight into an important population of partially obscured AGN that may constitute a significant fraction of high-redshift AGN. We present a preliminary sample of \silvernr\ candidates in the $\sim45$ arcmin$^{2}$ Abell 2744 field. Thanks to the large lensing area afforded by Abell 2744, the observations can reach up several magnitudes deeper than in blank fields.

Throughout, we assume a concordance cosmology with H$_0$=70, $\Omega_{\Lambda} = 0.7$, $\Omega_{M} = 0.3$ \citep{Hinshaw:2013}.

\section{Search for Red Compact Sources}
\label{sec:data}

\begin{figure*}
\includegraphics[width=0.3\textwidth]{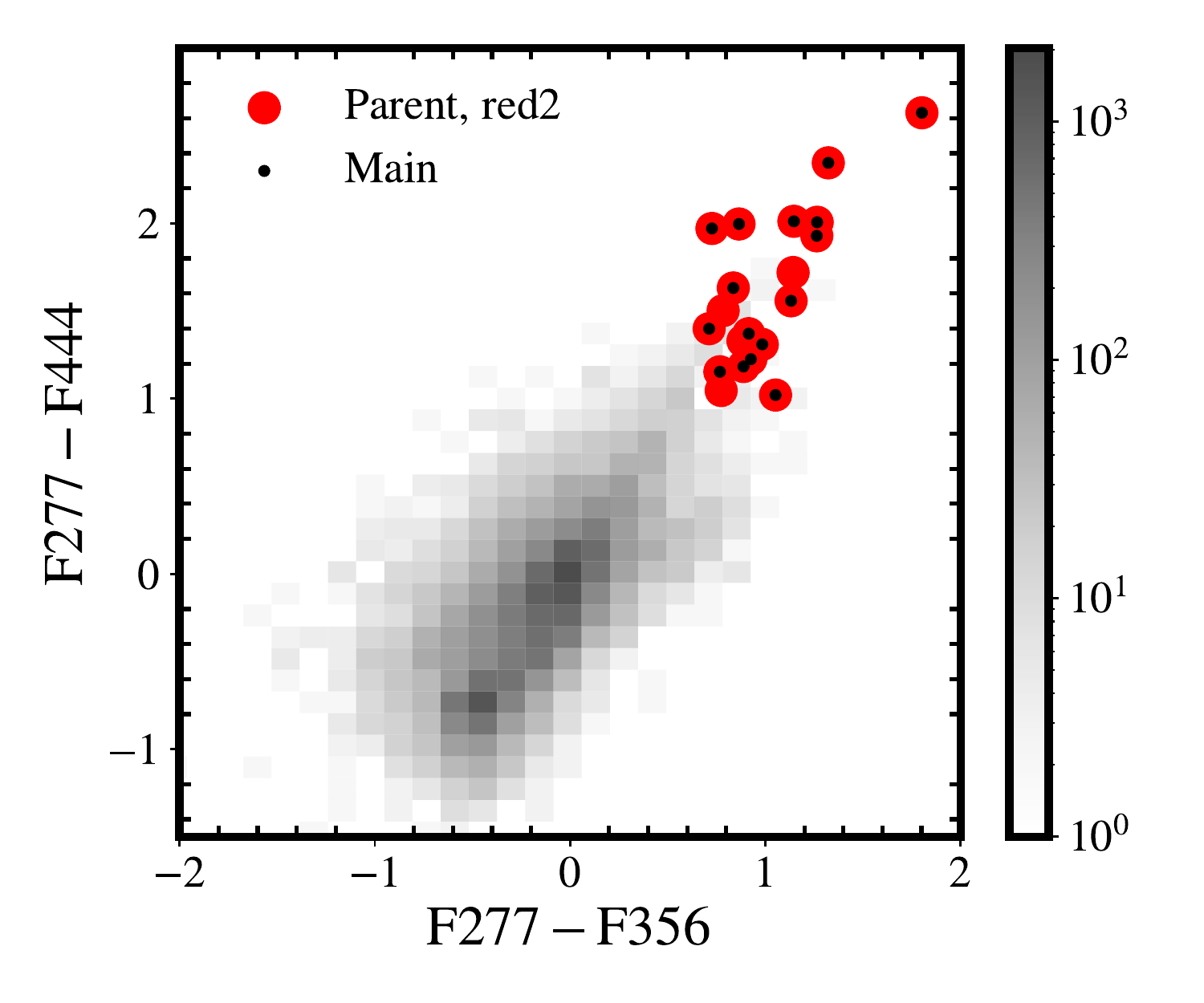}
\includegraphics[width=0.3\textwidth]{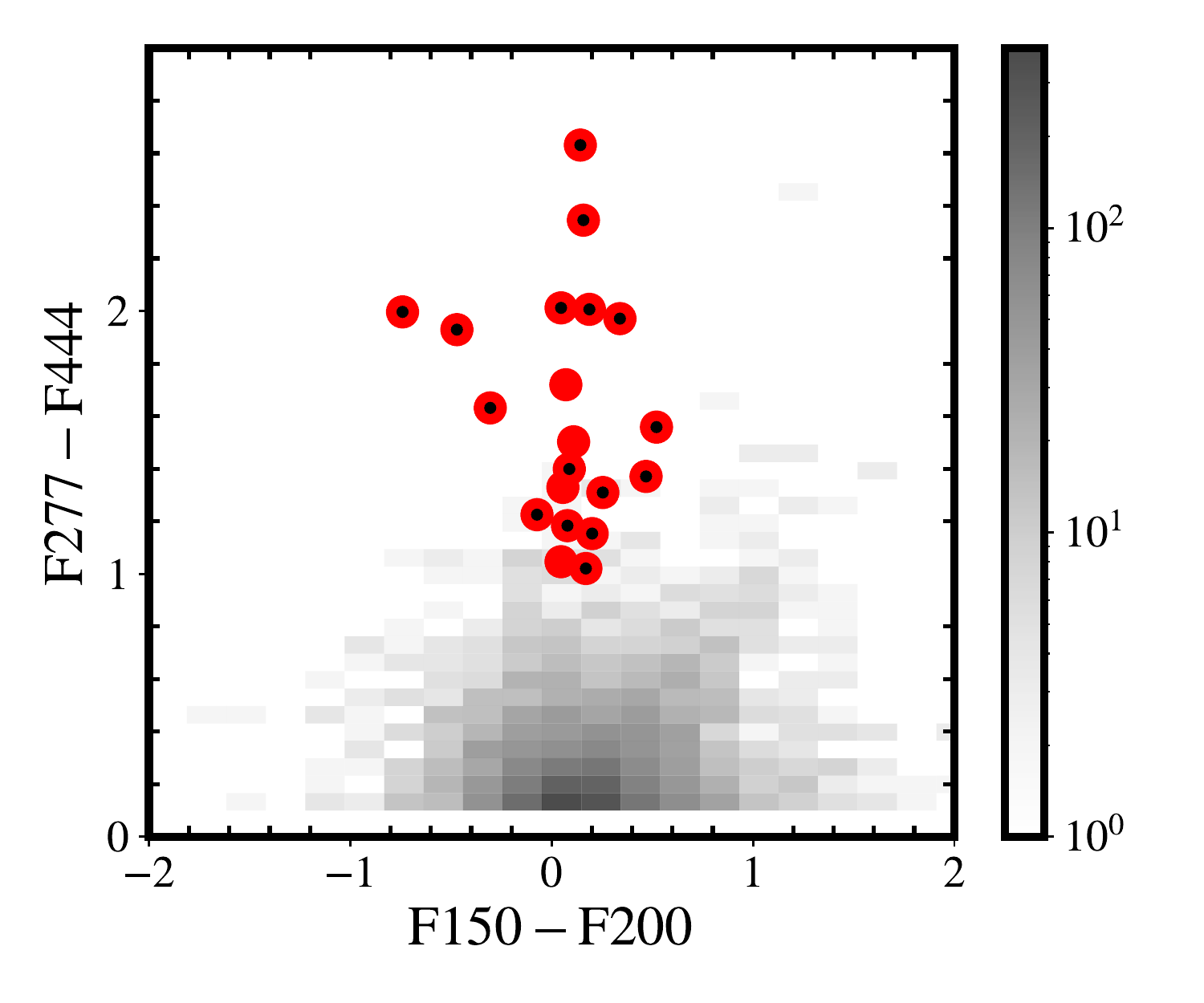}
\includegraphics[width=0.3\textwidth]{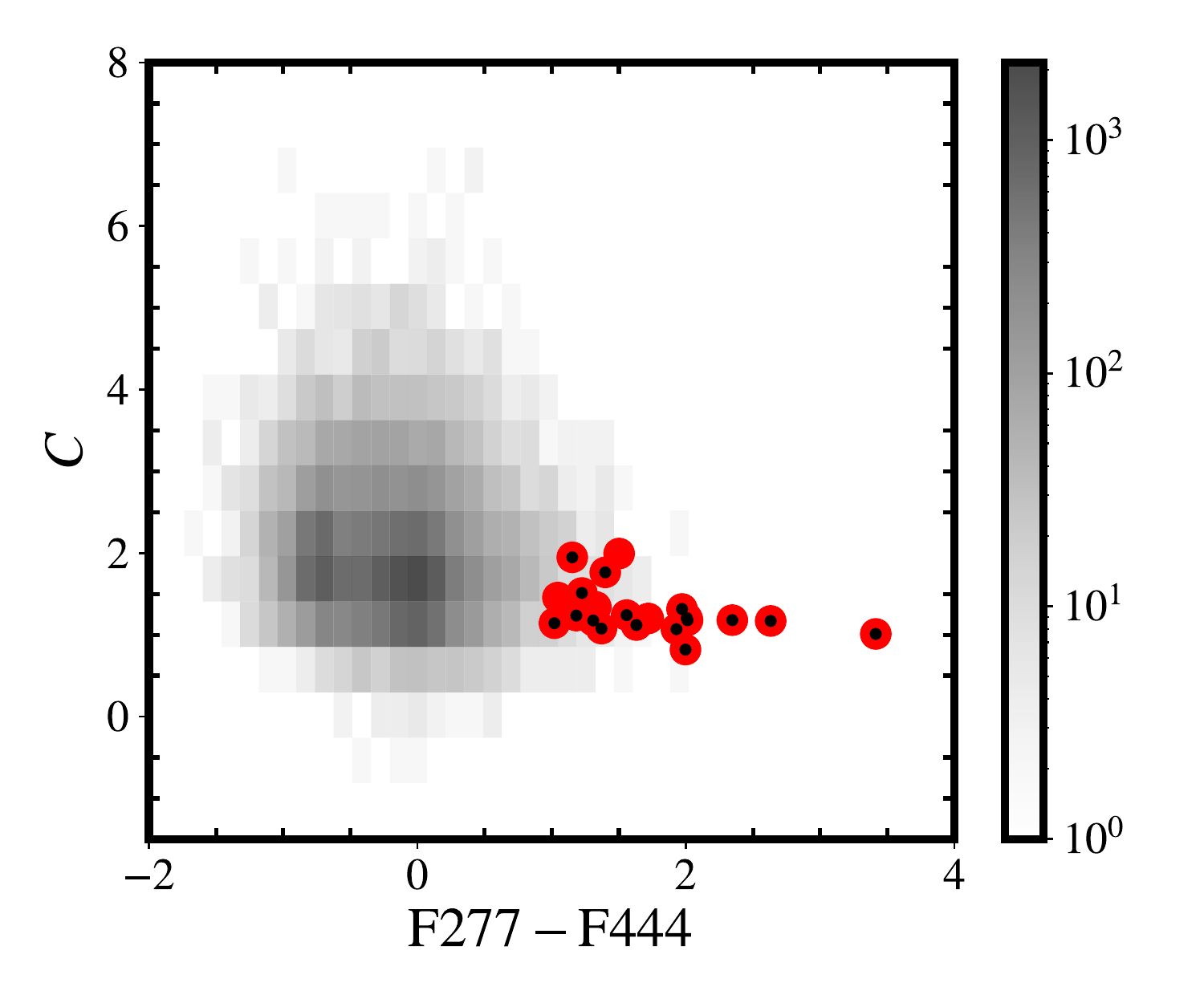}
\caption{Color-color (left) and color-compactness (right) selections. These panels show all the sources selected with the {\it ``red2''} color and compactness criteria (red circles), which is specifically designed to select $z>6$ AGN similar to those found in \citep{Furtak:2022,Kocevski:2023}. Sources selected to be PSF-dominated are highlighted with black dots. In greyscale, we show the entire UNCOVER catalog, with numbers of objects indicated by the color-bar. The compact red sources are clear outliers in color-color-compactness space. }
\label{fig:color-select}
\end{figure*}

The UNCOVER data are very well-suited to a search for more compact red sources. We have nearly complete wavelength coverage from $1-4~\micron$ to uniform depth, along with deep ALMA data to constrain the presence of cold dust \citep{Fujimoto:2023}. 

\subsection{The UNCOVER Survey}

Our search is performed using the \emph{JWST} Cycle 1 Treasury program Ultradeep NIRSpec and NIRCam ObserVations before the Epoch of Reionization \citep[UNCOVER;][]{Bezanson:2022}. UNCOVER imaging was completed in November 2022, comprising ultradeep ($29-30$~AB mag) imaging over 45 arcmin$^2$ in the galaxy cluster Abell 2744. This well-studied Frontier Field cluster \citep{Lotz:2017} at $z=0.308$ has one of the largest high-magnification areas of known clusters, and thus made an excellent target for deep (4-6 hours per filter) imaging across seven NIRCam filters (F115W, F150W, F200W, F277W, F356W, F410M, F444W). The nominal depth of $\sim 30$~mag can reach sources as faint as 31.5~mag with the help of magnification.  Photometric catalogs \citep{Weaver:2023} including existing \emph{HST} data have been made available to the public, and the lens model is also publicly available \citep{Furtak:2022b_SLmodel}. The initial selection of objects is based on the UNCOVER Data Release DR1 images and catalogs (2023). 

\citet{Fujimoto:2023} present deep ALMA 1.2~mm continuum imaging in Abell 2744. A wider, deeper 1.2~mm map of the full NIRCam UNCOVER area was newly obtained in Cycle 9 (\#2022.1.00073.S; S. Fujimoto in prep), 
reaching continuum r.m.s. sensitivity of $33 \mu Jy$ in the deepest areas. Prior-based photometry is extracted for all sources by measuring the ALMA flux in the natural resolution map (beam $\approx0.7-0.8\arcsec$) at the NIRCam positions.

In late July 2023, deep (2.7-19 hour) spectroscopic follow-up with the NIRSpec/PRISM will constitute the second phase of UNCOVER, but there are already many exciting things to find within the imaging \citep[e.g.,][]{Furtak:2023,Atek:2023}.

\subsection{Color and Morphology Selection}
\label{sec:firstcut}

\begin{deluxetable}{ccc}
\tabletypesize{\footnotesize}
\tablecolumns{3}
\tablewidth{0pt}
\tablecaption{ Sample Selection \label{table:samplesize}}
\tablehead{
\colhead{Sample} & \colhead{Cuts} & \colhead{Number}\\ }
\startdata
Parent & ($red1 \vert\ red2) \, \& \, compact$ & \bronzenr  \\ 
Main &  Dominated by PSF in 2D fit & \silvernr  \\ 
SED &  AGN-dominated at F444W in SED fits & \goldnr  \\ 
\enddata
\tablecomments{A simple flow chart of our selection. In the first step, we apply the combined color and compactness cut (\S \ref{sec:firstcut}) to select the {\it Parent} sample. Then we fit all sources in two dimensions and keep only those well-fit by a point source, which forms the {\it Main} sample (\S \ref{sec:psffit}). Finally, we generate an {\it SED} sample from those galaxies (\S \ref{sec:hiqual}).}
\end{deluxetable}

Inspired by the compact red broad-line AGN in CEERS, we devise a color+morphology search through the UNCOVER catalog for systems with both a moderate blue continuum and a rising red continuum, with the continuum break occurring between F150W and F356W for the typical source in our sample.  This break will move into F444W for $z \approx 8$, limiting the redshift range of the sample. The spirit of the selection is to find objects with a red continuum slope in the rest-frame optical, which also shows significant emission in the rest-UV. The color selection requires red colors in adjacent pairs of NIRCam bands to avoid selecting SEDs have blue continua throughout but with one band simply boosted by strong emission lines. We do not use photometric redshift or stellar population models, to avoid being limited by the models or templates used.

The initial Parent sample is selected with the following color selection. Starting with sources that are well-detected in the F444W band, with $SNR(444) > 14 \ \& \ m_{444} < 27.7$~mag, we select sources that are ({\it red1} $\vert$ {\it red2} ) \& {\it compact}, where 

$${\it red1} = (115-150 < 0.8) \,  $$
$$  (200-277 > 0.7) \, $$ 
$$ (200-356 > 1.0) \,  $$  

Or
$${\it red2} = (150-200 < 0.8) \,  $$ 
$$ (277-356 > 0.7) \,  $$
$$ (277-444 > 1.0)$$
And:
$$ {\it compact} = f_{444}(0.4\arcsec) / f_{444}(0.2\arcsec) < 1.7 $$. 

We find that extended dusty galaxies and inclined spirals over a wide range of redshift are contaminants when applying the color cuts alone. While such galaxies are certainly interesting \citep[e.g.,][]{Nelson:2023,Barrufet:2023}, we explicitly incorporate a compactness criterion, such that extended dusty galaxies are excluded and compact sources remain. The compactness criterion selects sources with a flux ratio up to $30\%$ higher than objects on the stellar locus. The precise limit does not matter and is mostly to reduce the sample size, as we will fit 2-D models to the Parent sample in \S \ref{sec:psffit} to search for sources that are likely dominated by a point source.

The UNCOVER survey is located in a gravitational lensing field, with a high density of bright galaxies with extended wings. This can cause the photometric aperture in the UNCOVER DR1 catalog to pick up light from nearby foreground sources and bias both the colors and the total magnitudes (the latter are based on a Kron-like autoscaling aperture). This is particularly an issue for the highest magnification sources \citep[see e.g.,][]{Furtak:2023}. To reduce the impact of contamination, we extract PSF-photometry by fitting 10-20 nearby isolated, bright, unsaturated stars to each source, fitting the position and flux in the F444W image, and then rescaling the flux in the other filters. This PSF photometry produces excellent agreement with the aperture photometry for isolated sources, with the benefit of strongly reduced contamination, especially in the bluer NIRCam SW filters, for sources with bright neighbors. For sources with additional complex extended structure the PSF-photometry is also more reflective of the nuclear component. It is clear from these initial fits that many sources are indistinguishable from a PSF (see e.g., appendix Figure \ref{fig:appendix_psf}).

From the initial color and compactness criteria, we select the Parent sample of \bronzenr\ AGN candidates out of a catalog of \catnr\ (see Table \ref{table:samplesize}). Next we will fit the sources with a point-source model to explore whether there is any evidence for extended light (\S \ref{sec:psffit}), and then examine the SEDs in detail (\S \ref{sec:SED}). The number of targets at each step is summarized in Table \ref{table:samplesize}. The basic sample properties are summarized in Table \ref{table:thesample}.

\begin{deluxetable*}{ccccccccccccc}
\tabletypesize{\footnotesize}
\tablecolumns{13}
\tablewidth{0pt}
\tablecaption{ Sample \label{table:thesample}}
\tablehead{
\colhead{ID} & \colhead{RA} & \colhead{Dec} & \colhead{$z_{\rm phot}$} & \colhead{$\mu$} & \colhead{Main} & \colhead{SED} & \colhead{F444W} & \colhead{277-356} & \colhead{277-444} & \colhead{$r_e$} & \colhead{F444W$_{\rm Sers}$} & \colhead{F444W$_{\rm PSF}$}  \\ 
\colhead{(1)} & \colhead{(2)} & \colhead{(3)} & \colhead{(4)} & \colhead{(5)} & \colhead{(6)} & \colhead{(7)} & \colhead{(8)} & \colhead{(9)} & \colhead{(10)} & \colhead{(11)} & \colhead{(12)} & \colhead{(13)}
}
\startdata
571 & 3.592423 & $-$30.432824 & 5.50 & 1.69 & 1 & 1 & 27.13 & 0.92 & 1.37 & 37 & 28.12 & 27.57 \\
 1967 & 3.619200 & $-$30.423270 & 6.01 & 1.63 & 1 & 0 & 24.67 & 1.15 & 2.01 & 3 & 26.03 & 25.27 \\
 2476 & 3.610205 & $-$30.421001 & 4.56 & 1.92 & 1 & 0 & 24.46 & 0.36 & 0.32 & 9 & 24.72 & 25.41 \\
 2940 & 3.575989 & $-$30.419030 & 4.47 & 1.81 & 1 & 1 & 27.02 & 0.98 & 1.31 & 32 & 27.79 & 27.75 \\
 3016 & 3.598798 & $-$30.418733 & 4.54 & 2.66 & 0 & 0 & 25.78 & 0.62 & 0.35 & 21 & 26.22 & 26.58 \\
 5957 & 3.556703 & $-$30.408192 & 3.52 & 1.57 & 1 & 1 & 24.72 & 0.48 & 0.78 & 3 & 23.58 & 24.63 \\
 6151 & 3.567758 & $-$30.407272 & 4.45 & 1.86 & 1 & 0 & 26.89 & 0.29 & 0.40 & 51 & 27.83 & 27.60 \\
 6430 & 3.550837 & $-$30.406598 & 4.98 & 1.54 & 1 & 1 & 24.43 & 0.44 & 1.07 & 9 & 26.63 & 24.54 \\
 8296 & 3.579829 & $-$30.401569 & 6.80 & 6.94 & 1 & 1 & 24.60 & 1.80 & 2.63 & 3 & 25.44 & 26.07 \\
 8602 & 3.612511 & $-$30.400595 & 6.51 & 1.86 & 0 & 0 & 25.54 & 0.77 & 1.05 & 62 & 25.88 & 27.04 \\
 8798 & 3.620605 & $-$30.399950 & 6.44 & 1.58 & 1 & 1 & 25.06 & 1.32 & 2.35 & 33 & 25.91 & 25.72 \\
 9992 & 3.583536 & $-$30.396676 & 6.79 & 8.55 & 1 & 1 & 25.06 & 0.71 & 1.40 & 22 & 26.42 & 25.40 \\
 10148 & 3.545794 & $-$30.395724 & 4.96 & 1.69 & 1 & 1 & 25.18 & 0.84 & 1.63 & 6 & 26.79 & 25.53 \\
 10712 & 3.597201 & $-$30.394328 & 6.77 & 4.12 & 1 & 1 & 26.36 & 1.27 & 2.01 & 26 & 26.98 & 26.97 \\
 13556 & 3.640410 & $-$30.386436 & 5.17 & 1.32 & 1 & 1 & 26.24 & 0.73 & 1.97 & 26 & 27.17 & 26.81 \\
 15798 & 3.535308 & $-$30.381010 & 6.44 & 2.54 & 1 & 0 & 27.43 & 1.05 & 1.02 & 25 & 27.74 & 29.14 \\
 16561 & 3.567022 & $-$30.379719 & 6.34 & 3.56 & 1 & 1 & 26.39 & 0.78 & 1.50 & 37 & 27.21 & 27.09 \\
 20080 & 3.569595 & $-$30.373224 & 5.52 & 2.71 & 1 & 1 & 26.81 & 1.26 & 1.93 & 26 & 28.31 & 27.11 \\
 21860 & 3.537529 & $-$30.370169 & 5.46 & 39.96 & 1 & 1 & 27.29 & 0.93 & 1.23 & 4 & 28.32 & 27.68 \\
 23778 & 3.546420 & $-$30.366245 & 7.02 & 4.05 & 1 & 1 & 27.09 & 0.87 & 2.00 & 20 & 27.67 & 28.05 \\
 28343 & 3.530008 & $-$30.358013 & 4.95 & 1.87 & 1 & 1 & 24.36 & 0.89 & 1.18 & 3 & 25.08 & 26.05 \\
 29466 & 3.513891 & $-$30.356024 & 7.22 & 1.54 & 1 & 1 & 25.74 & 1.37 & 3.41 & 2 & 26.38 & 26.41 \\
 30782 & 3.533997 & $-$30.353311 & 6.74 & 1.75 & 1 & 1 & 25.79 & 0.89 & 1.33 & 54 & 26.81 & 26.54 \\
 31142 & 3.498841 & $-$30.352945 & 4.81 & 1.00 & 1 & 0 & 23.12 & 0.86 & 0.68 & 2 & 24.22 & 24.45 \\
 31298 & 3.472870 & $-$30.352132 & 4.82 & 1.00 & 1 & 0 & 24.12 & 0.62 & 1.35 & 3 & 24.48 & 25.98 \\
 34061 & 3.506421 & $-$30.345969 & 6.64 & 1.38 & 0 & 0 & 24.99 & 1.14 & 1.72 & 42 & 25.39 & 26.34 \\
 35771 & 3.525296 & $-$30.342213 & 4.30 & 1.44 & 1 & 0 & 25.47 & 1.13 & 1.56 & 28 & 26.48 & 25.98 \\
 35819 & 3.473941 & $-$30.342061 & 4.59 & 1.00 & 1 & 0 & 26.98 & 0.56 & 0.50 & 12 & 29.18 & 27.08 \\
 35927 & 3.510632 & $-$30.341969 & 4.66 & 1.36 & 0 & 0 & 24.28 & 0.58 & 0.66 & 26 & 24.76 & 25.43 \\
 36867 & 3.465764 & $-$30.339833 & 4.78 & 1.00 & 0 & 0 & 26.33 & 0.43 & 0.11 & 29 & 26.46 & 28.01 \\
 37108 & 3.569464 & $-$30.339305 & 3.01 & 1.53 & 1 & 0 & 26.82 & 0.27 & 0.34 & 4 & 28.73 & 27.02 \\
 39955 & 3.475879 & $-$30.332051 & 3.64 & 1.00 & 0 & 0 & 26.36 & 0.20 & 1.13 & 28 & 26.74 & 27.65 \\
 46991 & 3.466191 & $-$30.315301 & 4.79 & 1.00 & 0 & 0 & 26.20 & 0.58 & 0.78 & 26 & 26.34 & 27.84 \\
 49555 & 3.470405 & $-$30.302487 & 3.40 & 1.00 & 0 & 0 & 26.48 & 0.28 & 0.19 & 400 & 26.90 & 28.49 \\
 49567 & 3.472454 & $-$30.302163 & 3.05 & 1.00 & 0 & 0 & 27.18 & 0.08 & 0.01 & 200 & 27.30 & 29.62 \\
\enddata
\tablecomments{Table of objects that satisfy $(red1 \vert red2)$ and $compact$. $a$ are the three images of the lensed compact red object presented in \citet{Furtak:2023}.
Column (1): UNCOVER ID. 
Column (2): R.A. Column (3): Dec.
Column (4): Photometric redshift $z_{\rm phot}$ from the AGN-only best-fit. 
Column (5): Magnification ($\mu$). 
Column (6): Flag for galaxies falling in the Main sample, consistent with being point sources (\S \ref{sec:psffit}).
Column (7): Flag for galaxies in the SED sample, with F444W emission dominated by AGN light (\S \ref{sec:SED}). 
Column (8): F444W mag.
Column (9): F277W-F356W color (mag).
Column (10): F277W-F444W color (mag). 
Column (11): $r_e$ (milliarcsec) from the single-component \sersic\ fit. 
Column (12): F444W magnitude from the \sersic\ part of the two-component PSF+\sersic\ fit. 
Column (13): F44W magnitude of the PSF component of the PSF+\sersic\ fit. The Main sample are selected to have F444W$_{\rm Sers} > $F444W$_{\rm PSF}$.}
\end{deluxetable*}

\begin{figure}
\includegraphics[width=0.45\textwidth]{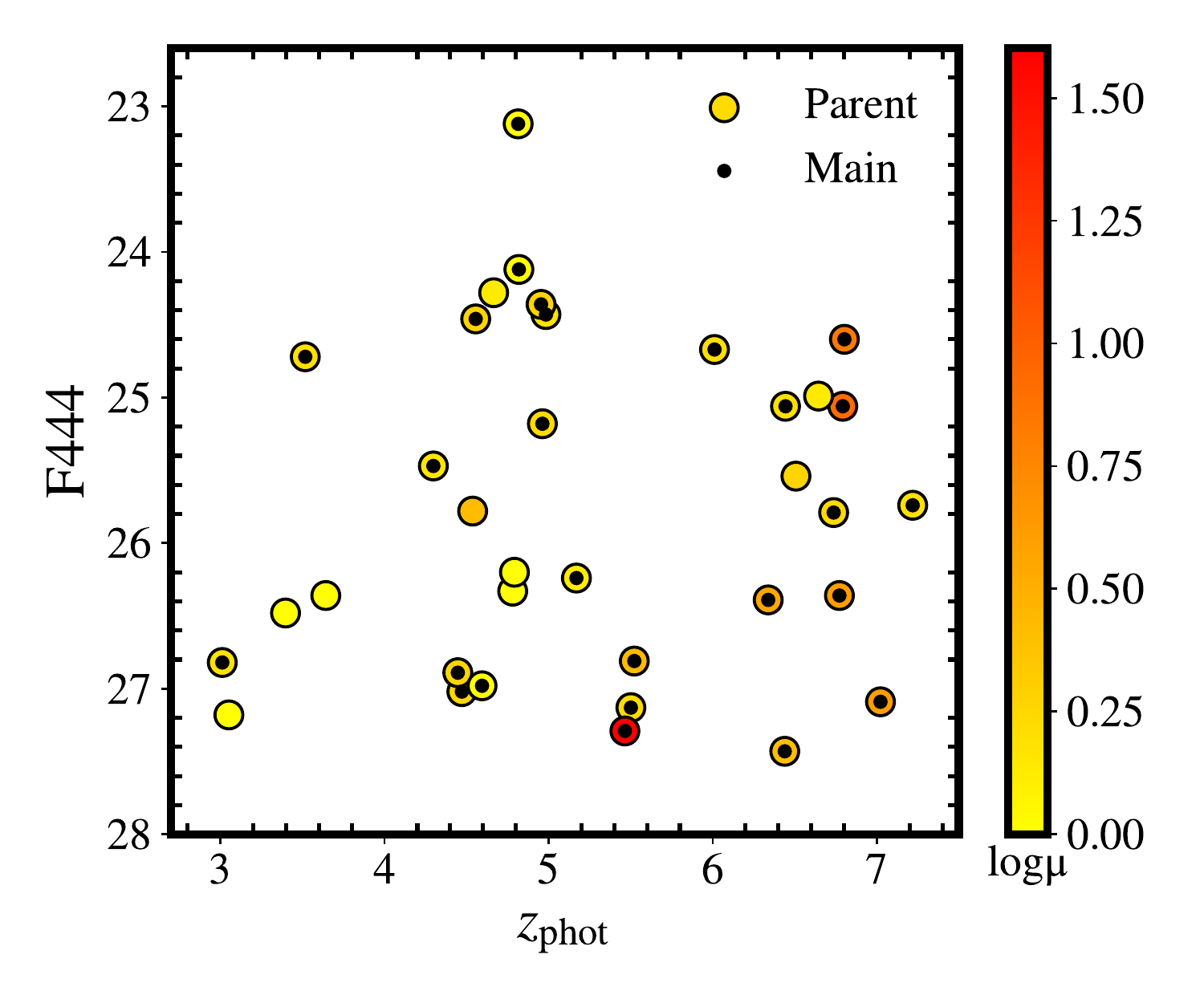}
\caption{The F444W magnitude and photometric redshift distributions of our sample, colored by the log of the magnification. Most objects in the sample have magnification $\sim 1.5$, but there are three strongly lensed systems, one of which \citet{Furtak:2023} has three images and so appears three times in this figure. We show the Parent sample, selected as $(red1 \vert red2) \, \& \, compact$, as well as the ``Main'' sample. The Main sample were selected to be point-source dominated, as described in \S \ref{sec:psffit}.}
\label{fig:summary}
\end{figure}

\subsection{Complementary Red Extended Sample}
\label{sec:DSFG}

If we select all galaxies that satisfy ({\it red1} $\vert$ {\it red2}) but have $compact > 1.7$, we have a sample of 16 galaxies. These galaxies span a wide range of redshift $2<z<6$ and appear typically powered by dusty star-formation taking place in an extended region, and they are nearly all detected in ALMA. Below, in \S \ref{sec:SED}, we will contrast their far-infrared-to-mm SEDs with the compact red sources as one additional argument that we are likely identifying sources with hotter dust than expected from dusty star formation.

\subsection{Joint PSF+\sersic\ fits}
\label{sec:psffit}

To refine the compactness criteria, we  perform a series of two-dimensional fits to the light profile of each galaxy in F444W using {\tt Galfit} \citep{Peng:2002,Peng:2010}. The main purpose is to determine if a source is unresolved, or alternatively dominated by a point-source component, taking into account variations in the PSF with position and magnitude. We choose to model the sizes in the F444W filter. While the resolution is higher in NIRCam SW, the sources have much higher SNR ratio in F444W owing to the very red colors. In addition, the size at the longest wavelength is less affected by dust and thus closer to the true distribution of luminosity or stellar mass. Bright, isolated, unsaturated stars are used as PSF models, as it is known that the simulated PSFs from \texttt{WebbPSF} \citep{Perrin:2014} are too narrow compared to empirical PSFs \citet[e.g.,][]{Ding:2022,Weaver:2023}. Stars are selected based on a cut in flux ratio between $D=0.2"$ and $D=0.4"$ apertures and an inspection of SEDs to retain sources with stellar SEDs. There are 10-20 nearby suitable PSF stars for each source. Before fitting, a local background is subtracted at scales of $40$ pixels ($1.6\arcsec$) by growing all segmentation maps by 2 pixels and subtracting the median of the background pixels.

\begin{figure*}
\hskip +10mm
\includegraphics[width=0.90\textwidth]{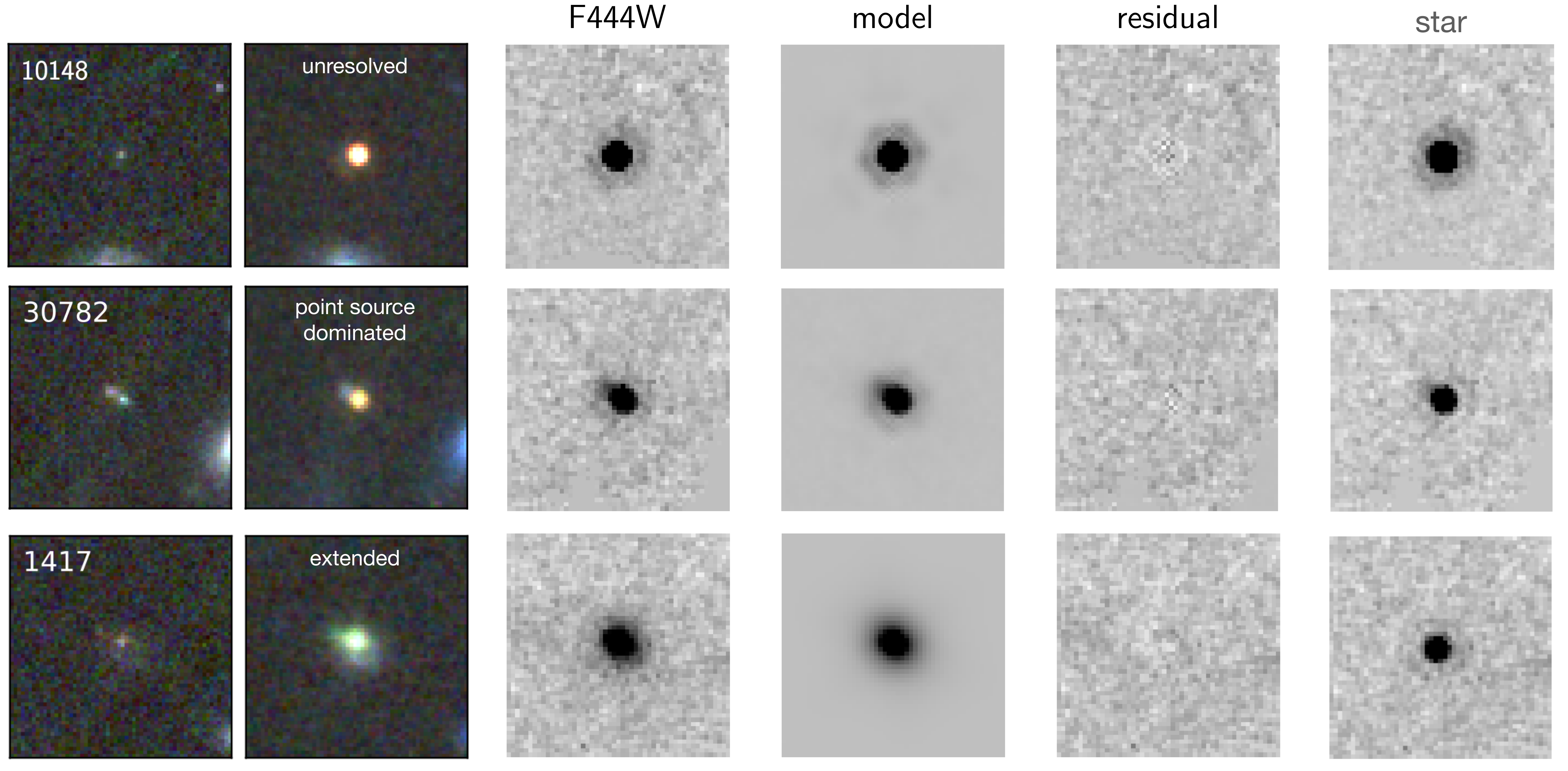}
\caption{Examples of Sersic profile fits to the F444W image. From left to right: NIRCam SW color image (based on F115W, F150W, F200W), NIRCam LW color image (F277W,F356W,F444W), F444W image, GALFIT best-fit model, residual, and a star scaled to the same flux placed in the residual map. The top row shows an unresolved source modeled with a single Sersic component that is indistinguishable from a star, the middle row shows a source that is better modeled by a two-component PSF+Sersic model, but is dominated by the PSF component. The bottom row shows an extended source without evidence for a bright point source. These are removed from the sample.}
\label{fig:twodPSF}
\end{figure*}

\begin{figure}
\includegraphics[width=0.45\textwidth]{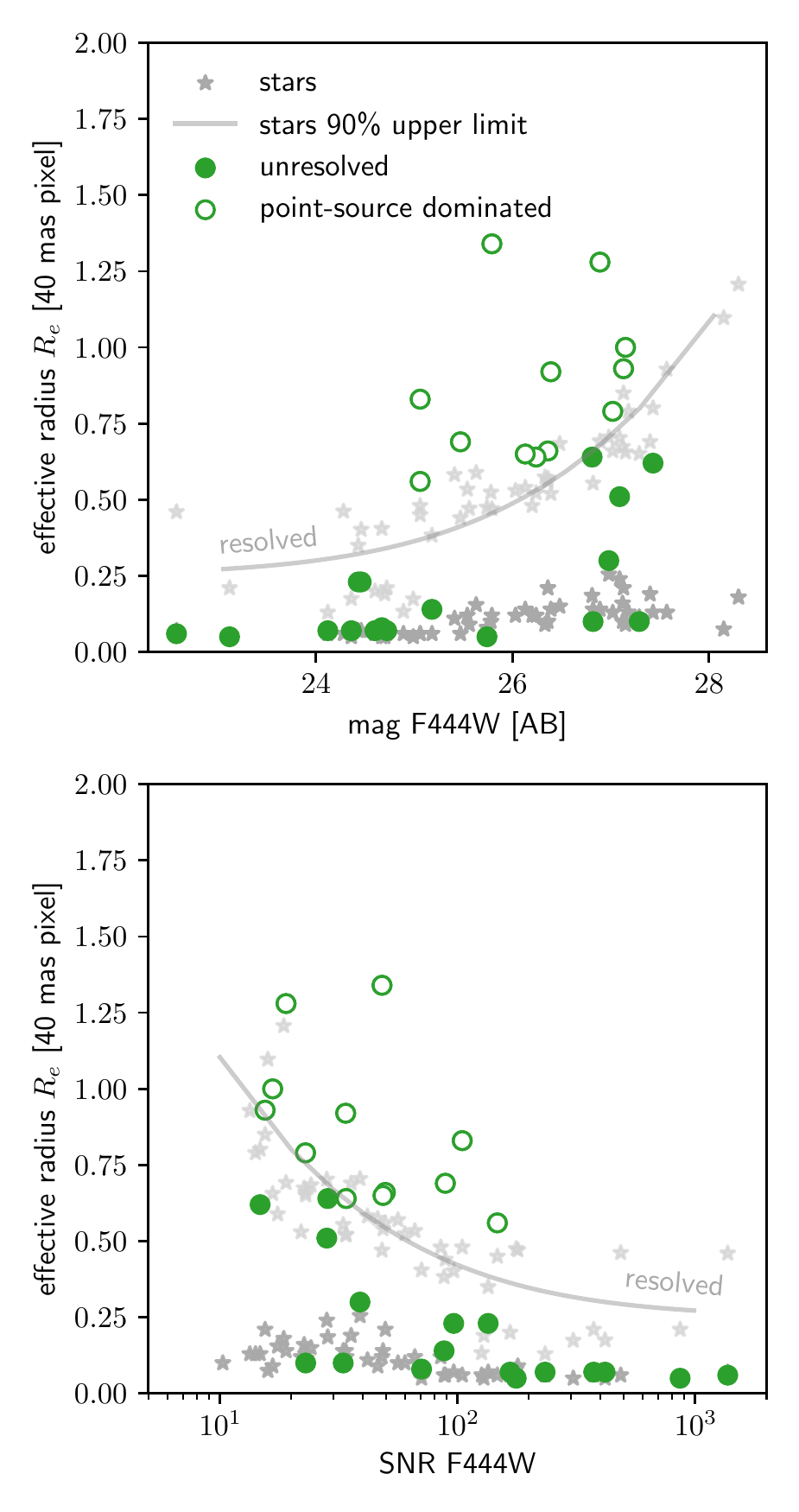}
\caption{The effective radii $R_e$ of the sample (green circles), compared to the effective radii measured on stars scaled to the same SNR, to determine if a source is resolved or consistent with being dominated by a point source. The best-fit $R_e$ of stars are shown by dark gray symbols with the $90\%$ upper envelope based on bootstrap resampling shown in light gray. Sources are considered point-source dominated if the $R_e$ falls below the 90\% envelope for stars (solid circles), or if a two-component PSF + Sersic fit indicates most of the light is in a point source (open circles). } 
\label{fig:sizedist}
\end{figure}

We start with single-component \sersic\ fits. We fit each galaxy using one of the nearby stars as a PSF model. Uncertainties are determined by taking the best-fit model, placing it in the residual map of another randomly drawn source in the sample, and re-fitting the source with a randomly drawn nearby PSF star. This process is repeated 200 times using unique combinations of PSFs and residuals, reflecting both systematic and random uncertainties. To determine whether the sources are consistent with point sources, the same process was performed on stars. For each source a random nearby star was drawn, scaled to the same SNR, placed in the residual map, and then modeled with a single \sersic\ model using another nearby star as the PSF. This process was repeated 200 times as well, using different combinations of stars, PSFs, and residual maps. We then examine the median and upper 90\% size of the fits as a function of their F444W magnitude and signal-to-noise ratio (SNR; Figure \ref{fig:sizedist}). Any source that is smaller than the 90\% percentile of the stellar fits is considered consistent with a PSF and therefore unresolved. This is the case for 17 out of the original \bronzenr\ sources. 

\begin{figure*}
\hspace{15mm}
\includegraphics[width=0.74\textwidth]{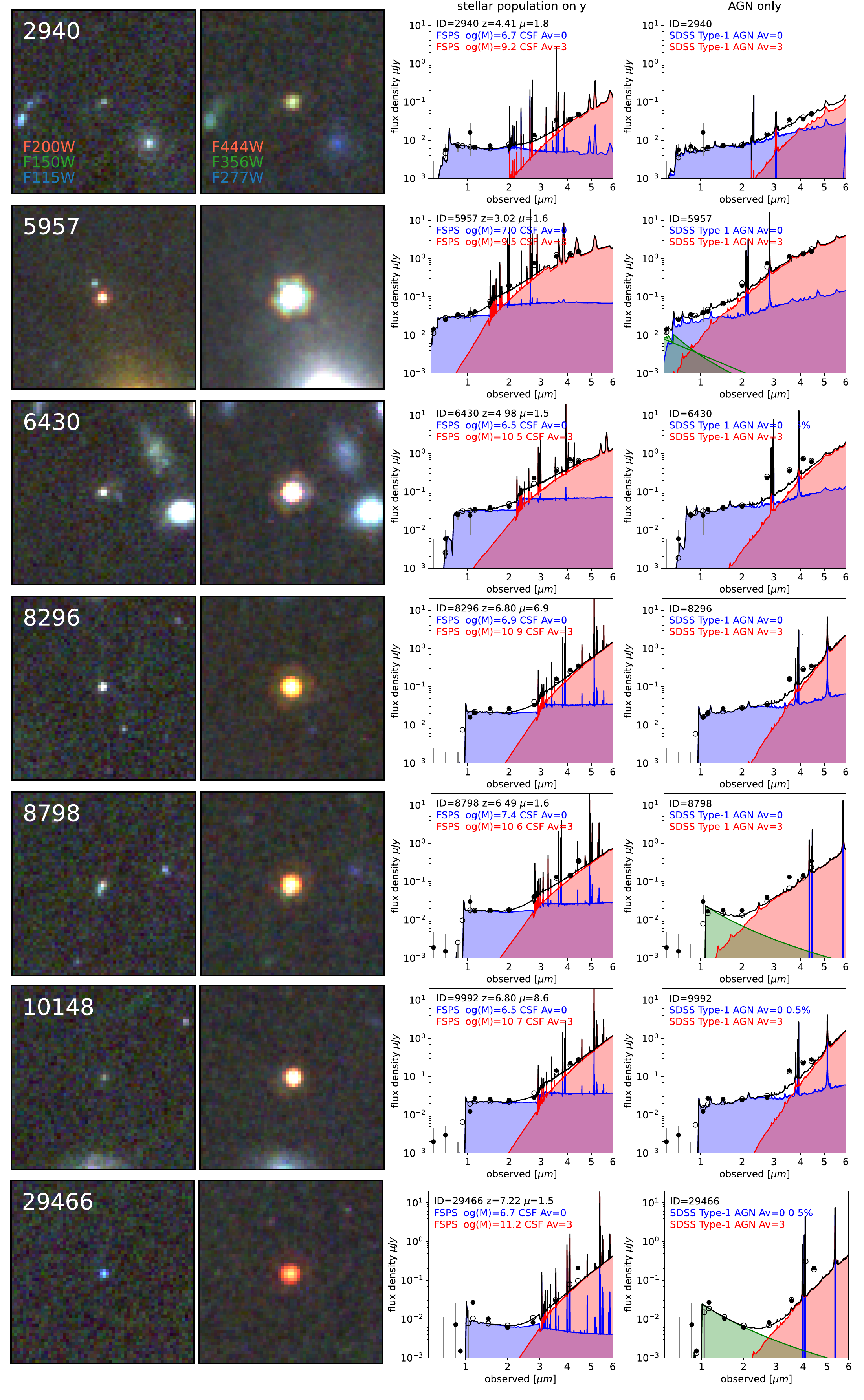}
\caption{Color image stamps and SED fits to the SED sample of compact red sources (see details in \S \ref{sec:hiqual}). The color coding is based on the NIRCam imaging in F115W (blue), F150W (green), F200W (red) in the left column and F277W, F356W, F444W in the right column. The images convey how compact the sources are, and that there is both a prominent red and prominent blue component to the SED. The SED fits are in all cases composite. In the third columns, we show the galaxy-only fits, while in the fourth column are displayed AGN-only fits. The data points are shown in filled black, while the model photometry are in open circles. Total model is in black, while reddened and unreddened components are in blue and red respectively. The green component is an additional blue power-law that is required to fit the SEDs of 6 of the SED-selected targets. Details of the fitting are discussed in \S \ref{sec:SED}.}
\label{fig:phot}
\end{figure*}

Some, however, show more complex morphology with extended structure in addition to a bright point-like source, as might be expected for an AGN embedded in a host galaxy. We then perform a two-component PSF+\sersic\ fit. From the two-component fits, we identify an additional 9 sources that are dominated by a PSF component, where more than half the light comes from the point-source component. This leaves a total of \silvernr\ sources that the two-dimensional fits deem to be point-source dominated. Of the \bronzenr\ in the Parent sample, \silvernr\ targets are PSF-dominated based on our joint PSF+\sersic\ fits -- the ``Main'' sample. See Figure \ref{fig:phot} for images that demonstrate the compact nature of these red sources.

We can translate our two-dimensional fits into size limits for the sample. At the magnitude limit F444W=$27.5$ AB of our sample, we can measure sizes to a bit under one pixel (Figure \ref{fig:sizedist}), or $R_e\lesssim0.03\arcsec$ at $90\%$ confidence. At the bright end sizes can be measured to better than $R_e\approx0.01\arcsec$. At the median distance of the objects redshift $z\sim5$ these limits correspond to $\sim140$~pc and $\sim40$~pc respectively, where a simple correction by $\mu^{-1/2}$ is applied to correct for lensing. The median size of the sample is $\langle r_e\rangle=54_{-10}^{+33}$~pc, with the uncertainties derived from bootstrap resampling. We note that the majority of the sample are only moderately lensed -- 24/26 have $\mu<3$ and are far away from caustics with low shear and modest relative uncertainties on the magnification (Figure \ref{fig:summary}). In the two cases of highly magnified sources (ID=8296, 21860), the size limits are considerably tighter \citep{Furtak:2023} but more detailed modeling taking into account the lens model uncertainties and shear is required.

It is tempting to assume that these small sizes point to an AGN origin for the emission. However, very compact star-forming galaxies are known to exist at lower redshift \citep[e.g.,][]{Geach:2018} and we do not have strong constraints on the mass-size relation for massive star-forming galaxies at $4<z<7$. Some very compact massive galaxies have been found in early \emph{JWST} data \citep[e.g.,][]{Carnall:2023,Robertson:2023,Morishita:2023,Williams:2023}, and extremely small sizes $<R_e>\approx150$~pc are reported for candidate massive galaxies $7 < z < 9$ \citep{Baggen:2023}. Apart from a strong triply-lensed system with size limits $<30$~pc \citep{Furtak:2023} (also in this sample, with ID=8296, 9992, 10712), the morphologies alone cannot distinguish between a stellar and accretion origin for the bulk of the sources. In the following section, we will present SED fits, and further isolate the ``SED'' sample of \goldnr\ objects that are likely to be AGN-dominated based modeling of the spectral energy distributions (SEDs).

\section{Spectral Energy Distributions and Number Densities of the Compact Red Sources}
In this section, we model the spectral properties of the new sample of compact red sources. We fit the seven NIRCam UNCOVER bands, along with \emph{HST} photometry and ALMA 1.2~mm data where available. By selection, the sources have blue continua in the NIRCam SW or rest-frame UV, typically $\beta_{\rm UV} \approx-2$, and rising red slopes in NIRCam LW filters corresponding to the rest-frame optical, with $\beta_{\rm opt} \approx 2, f_{\lambda} \propto \lambda^{\beta_{\rm opt}}$  (e.g., Figure \ref{fig:ceers3210-phot}). There is a clear inflection or break between F150W and F356W depending both on the intrinsic $A_V$ and on the redshift. Such composite SEDs can be challenging to model with any single galaxy or AGN model. Because of their composite color components, we model the SEDs as composites as well, with either multiple galaxy components (young, dusty star-forming, and evolved stellar populations), or with AGN components included.

It is useful to examine one example in a bit more detail \citep[and see also discussion in][]{Barro:2023}. \citet{Furtak:2022} presented a highly magnified compact red source with three images in the UNCOVER field. Because of the high magnification, this source has a particularly stringent limit on its size of $r_e \sim 35$~pc, making stellar origins for the emission seem even less likely than the typical source in our sample. Furtak et al.\ present three possible galaxy models to fit the SED. Prospector$-\beta$ \citep{Wang:2023} prefers to fit the red color with a massive evolved galaxy and strong Balmer break, but cannot simultaneously fit the UV emission. EAZY \citep{Brammer:2008} is well suited to handle different components with different reddening, but still is unable to satisfactorily fit the strong red upturn. Modeling software that includes optical/UV AGN emission also struggles to find multi-component solutions: \citeauthor{Furtak:2022} present fits from both CIGALE \citep{Boquien:2019} and Beagle+AGN \citep{Chevallard:2016,Vidal-Garcia:2022}. The latter only includes narrow emission lines from an AGN and thus attempts to reproduce the red NIRCam LW colors by extreme narrow-emission lines to boost the fluxes in the broad band filters. This seems unlikely to arise from a 35~pc radius and the need to boost the NIRCam filters by a factor $\times10$ leads to extreme equivalent widths $>10,000$~\AA. In the context of the compact red sources in CEERS that have broad H$\alpha$ in their spectra at much lower equivalent widths ($\sim400$\AA), the continuum is arguably AGN-dominated, with the steep red continuum due to a reddened broad-line AGN continuum. The UV component might arise from scattered light \citep[e.g.,][]{Assef:2020,Glikman:2023}, patchy transmission, or star formation. That possibility is shown in the AGN-only fits in Figure \ref{fig:phot}. 

We have thus performed custom fitting to handle our need for multiple components that each may have different reddening and AGN contributions. We will describe the general properties of the fits in this section, but refer to the Appendix for more details.

\begin{deluxetable*}{crcccccccccc}
\tabletypesize{\footnotesize}
\tablecolumns{12}
\tablewidth{0pt}
\tablecaption{ SED Fits \label{table:thefits}}
\tablehead{
\colhead{ID} & \colhead{$f_{\rm ALMA}$} & \colhead{$e_{\rm ALMA}$} & \colhead{$f_{\rm AGN}$} & \colhead{$M_{\rm *,gal+ALMA}$} & \colhead{$M_{\rm *, noALMA}$}  & \colhead{log ($L_{\rm bol}$)} & \colhead{$M_{1450}$} & \colhead{$\chi^2_{\rm AGN}$} & \colhead{$\chi^2_{\rm gal,noALMA}$} & \colhead{$\chi^2_{\rm gal+ALMA}$} & \colhead{SED}  \\ 
\colhead{(1)} & \colhead{(2)} & \colhead{(3)} & \colhead{(4)} & \colhead{(5)} & \colhead{(6)} & \colhead{(7)} & \colhead{(8)} & \colhead{(9)} & \colhead{(10)} & \colhead{(11)} & \colhead{(12)} 
}
\startdata
571 & 0.48 & 4.02 & 3.61 & 9.0$\pm$0.1 & 10.1$\pm$0.2 & 44.4 & $-$17.6 & 11.6 & 28.8 & 98.0 & 1 \\
 1967 & \nodata & \nodata & 0.00 & 10.9$\pm$0.1 & 10.9$\pm$0.1 & 45.3 & $-$18.3 & 12.0 & 17.4 & 10.7 & 0 \\
 2476 & $-$4.31 & 4.78 & 0.86 & 10.0$\pm$0.1 & 10.0$\pm$0.1 & 45.1 & $-$19.5 & 24.8 & 49.5 & 47.8 & 0 \\
 2940 & $-$3.55 & 3.94 & 6.72 & 9.0$\pm$0.1 & 9.5$\pm$0.3 & 43.8 & $-$16.4 & 5.7 & 26.3 & 32.4 & 1 \\
 5957 & $-$0.97 & 4.21 & 42.56 & 10.4$\pm$0.1 & 10.6$\pm$0.1 & 45.0 & $-$17.5 & 38.8 & 37.5 & 68.7 & 1 \\
 6151 & 0.39 & 3.93 & 0.00 & 8.7$\pm$0.1 & 9.1$\pm$0.1 & 43.8 & $-$16.7 & 50.1 & 84.8 & 90.6 & 0 \\
 6430 & $-$4.35 & 6.83 & 2.90 & 10.1$\pm$0.1 & 10.5$\pm$0.2 & 45.5 & $-$18.1 & 7.9 & 51.1 & 237.4 & 1 \\
 8296 & 0.26 & 3.92 & 5.99 & 8.8$\pm$0.1 & 11.0$\pm$0.1 & 46.2 & $-$16.5 & 12.2 & 36.8 & 375.8 & 1 \\
 8798 & 1.89 & 4.33 & 14.02 & 9.6$\pm$0.1 & 10.5$\pm$0.1 & 45.5 & $-$18.0 & 23.6 & 6.2 & 264.6 & 1 \\
 9992 & 0.50 & 3.90 & 5.32 & 8.9$\pm$0.1 & 10.9$\pm$0.1 & 45.9 & $-$16.5 & 38.7 & 54.7 & 305.3 & 1 \\
 10148 & 5.15 & 4.23 & 12.38 & 9.8$\pm$0.1 & 10.5$\pm$0.1 & 45.3 & $-$16.9 & 15.7 & 65.5 & 333.1 & 1 \\
 10712 & 2.24 & 3.34 & 11.93 & 9.1$\pm$0.1 & 10.6$\pm$0.2 & 45.7 & $-$16.0 & 20.7 & 48.4 & 277.5 & 1 \\
 13556 & $-$3.94 & 4.38 & 5.66 & 9.0$\pm$0.1 & 9.4$\pm$0.3 & 45.2 & $-$16.9 & 15.0 & 16.1 & 111.6 & 1 \\
 15798 & 1.45 & 4.22 & 0.00 & 8.0$\pm$0.2 & 8.4$\pm$0.2 & 43.8 & $-$16.8 & \nodata & 33.7 & 64.2 & 0 \\
 16561 & $-$3.76 & 4.18 & 2.25 & 8.9$\pm$0.1 & 9.7$\pm$0.3 & 44.7 & $-$16.4 & 17.5 & 27.5 & 71.2 & 1 \\
 20080 & $-$3.06 & 3.40 & 13.23 & 9.2$\pm$0.1 & 10.2$\pm$0.2 & 44.2 & $-$16.9 & 22.9 & 37.6 & 197.8 & 1 \\
 21860 & 1.93 & 4.21 & 2.19 & 7.5$\pm$0.1 & 8.5$\pm$0.2 & 42.9 & $-$14.5 & 64.4 & 59.9 & 119.3 & 1 \\
 23778 & $-$1.27 & 4.20 & 4.89 & 9.0$\pm$0.1 & 9.6$\pm$0.3 & 45.3 & $-$18.0 & 29.7 & 446.6 & 673.2 & 1 \\
 28343 & 0.72 & 4.21 & 7.37 & 9.8$\pm$0.1 & 9.7$\pm$0.2 & 44.9 & $-$18.7 & 7.7 & 57.4 & 456.4 & 1 \\
 29466 & \nodata & \nodata & 25.31 & 11.4$\pm$0.1 & 11.4$\pm$0.1 & 46.2 & $-$18.2 & 61.9 & 340.1 & \nodata & 1 \\
 30782 & 2.11 & 3.98 & 4.84 & 9.4$\pm$0.1 & 10.4$\pm$0.1 & 45.3 & $-$17.9 & 6.6 & 13.1 & 149.7 & 1 \\
 31142 & \nodata & \nodata & 0.00 & 10.2$\pm$0.1 & 10.1$\pm$0.1 & 45.6 & $-$19.7 & 66.6 & 12.3 & \nodata & 0 \\
 31298 & \nodata & \nodata & 0.00 & 11.0$\pm$0.2 & 10.9$\pm$0.2 & 45.9 & $-$18.7 & 93.1 & 120.0 & \nodata & 0 \\
 35771 & \nodata & \nodata & 0.00 & 10.1$\pm$0.2 & 10.0$\pm$0.3 & 44.6 & $-$17.2 & 19.3 & 11.7 & \nodata & 0 \\
 35819 & \nodata & \nodata & 0.00 & 8.9$\pm$0.1 & 8.9$\pm$0.1 & 44.0 & $-$17.0 & 15.9 & 10.4 & \nodata & 0 \\
 37108 & 2.86 & 3.96 & 0.00 & 9.1$\pm$0.1 & 9.0$\pm$0.0 & 43.1 & $-$16.2 & \nodata & 10.6 & 18.6 & 0 \\
\enddata
\tablecomments{Column (1): Galaxy ID. Column (2): ALMA 1.2~mm flux based on forced photometry (units of 10~$\mu$Jy). Column (3): ALMA flux errors (10~$\mu$Jy). Column (4): AGN fraction from the combined fit at F444W. Column (5): Stellar masses from galaxy-only fits including ALMA ($M_{\odot}$). Column (6): Stellar masses from galaxy-only fits excluding ALMA ($M_{\odot}$). Column (7): Bolometric luminosity (erg/s) inferred from AGN-only fits, magnification-corrected and assuming that $L_{\rm bol} = 10 L_{5100}$. Column (8): $M_{1450}$ (mag) magnification-corrected. Column (9): $\chi^2$ of the best-fit AGN-only model including ALMA. Column (10): The $\chi^2$ for the best-fit galaxy-only model, without ALMA. Column (11): The $\chi^2$ for the best-fit galaxy-only model including ALMA. In cases with strong ALMA constraints, the $\chi^2$ is dramatically larger due to the inclusion of the ALMA upper limit. Column (12): Flag for galaxies in the SED-selected sample.} 
\end{deluxetable*}

\subsection{Composite model fits}
\label{sec:SED}

In the AGN-only fits we model the SED with a two-component AGN model. Each AGN component is comprised of the \citet{VandenBerk:2001} AGN template in the optical/UV, combined with the template from \citet{Glikman:2006} extending to the rest-frame near-infrared. In addition to the unreddened AGN, we also include a reddened version of the same template, with reddening ranging from $A_V = 0.5-5$ applied using the \citet{Calzetti:2001} reddening law. The reddened component is the primary component in terms of luminosity and must be heavily extincted to fit the red slope. The blue component is unreddened but weak, and we interpret this as scattered AGN light at a few percent of the luminosity of the reddened component. This component is generally well-fit by an unreddened AGN template, as expected for electron scattering, but we highlight some outliers at the end of this section. In some cases we may well be seeing star formation from the host; we note that in a small number of objects, the rest-UV emission does appear slightly extended (Figure \ref{fig:phot}). 

When our fits include the ALMA 1.2~mm point, we model the mid-to-far infrared emission with the CLUMPY torus models \citep{Nenkova:2008mod,Nenkova:2008obs}. Details are in the Appendix \citep[see also][]{Conroy:2009,Leja:2018}, but note that we fix the clump properties that determine the shape of the torus model and we fix the inclination to 40 degrees, such that we only incorporate mid-to-far infrared emission and not rest-frame UV/optical emission from the accretion disk. Also key is that we scale the infrared emission to the obscured optical luminosity, thus enforcing energy balance. More detailed constraints on the torus parameters would require longer-wavelength data. In addition, because we are using fixed empirical AGN templates, we do allow for small differences in the emission-line strengths, applying a Gaussian prior with a width of 0.3 dex. The goodness of fit for the AGN-only models is included in Table \ref{table:thefits}. Finally, in a minority of cases (six of the \silvernr), the UV continuum is steeper than the AGN or star-forming galaxy template, and in these cases we additionally include a UV power-law component. Thus, we are fitting a total of eight free parameters including the redshift. The details of the fitting routine are described in Appendix A. 

The galaxy-only fits are performed in the same spirit, but rather than AGN templates we include star-forming templates with a range of $A_V=0-5$ \citep{Calzetti:2001}, and a quiescent galaxy component. When we jointly model the ALMA data, we utilize the dust emission models as implemented in FSPS \citep[see description in][]{Leja:2017}. As with the infrared emission from the AGN, we fix the parameters that set the shape of the dust emission SED since we have minimal long-wavelength constraints. We also enforce energy balance between the reddened UV/optical continuum and the dust emission.


The photometric redshift is a free parameter in our AGN-only fits. We find decent agreement between our $z_{\rm phot}$ and the catalog value derived from EAZY \citep{Brammer:2008}, with a median $\langle \delta z /(1+z) \rangle = 0.04$ and scatter $\sigma z / (1+z)\approx 0.15$. There is a minority of cases where the AGN-only fits return a redshift $\delta z \approx 1-2$ lower than EAZY, due to a degeneracy between a featureless red slope (in the case of a reddened AGN) and a strong Balmer/4000\AA\ break \citep[CEERS 3210 is one example; ][]{Labbe:2023,Kocevski:2023}, in particular when the optical constraints are weak. When we have deep optical photometry from \emph{HST}, it can be possible to rule out the higher-z solution through the presence of a Lyman Break. Overall, the different fits yield reasonable agreement on their photometric redshifts.

In Figure \ref{fig:phot} (third column) we show example galaxy-only fits. To model the SEDs typically requires an unreddened (blue) star-forming galaxy template and a star-forming but heavily dust-obscured component. In a small number of cases, there are sources that can be fit by a combination of an evolved population with a strong break and a star-forming component. Indeed, there is at least one spectroscopic example of a galaxy-AGN composite with a strong break and a broad H$\alpha$ line \citep{Carnall:2023}, but in general our modeling does not find strong evidence of 4000\AA\ breaks, although in principle dust-reddened galaxies with strong breaks could also fit the observed SEDs \citep[see also][]{Kocevski:2023,Chworowsky:2023,Barro:2023}. Rather, the very red continuum slope is better fit by dusty star-formation. 

\begin{figure*}
\includegraphics[width=0.99\textwidth]{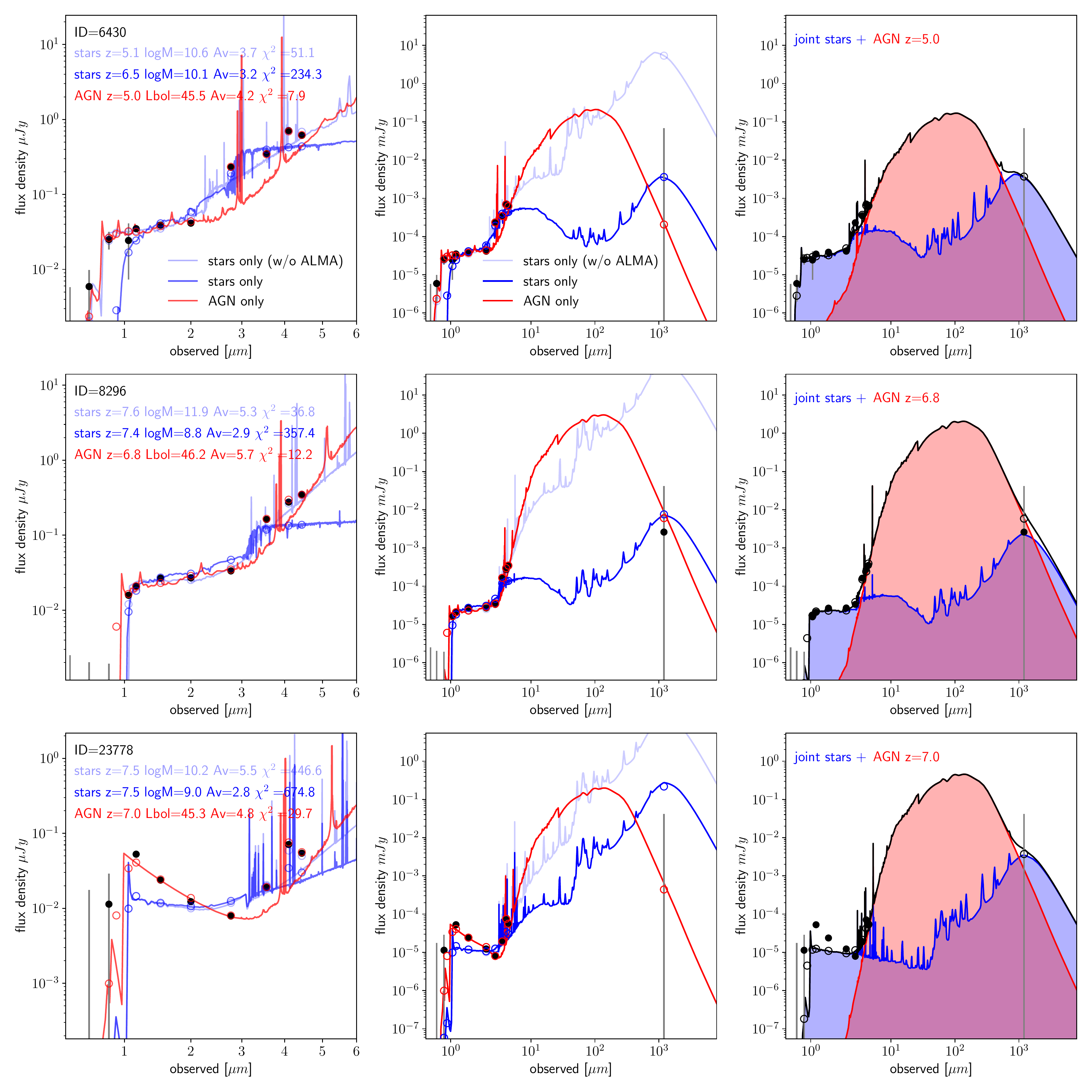}
\caption{Three examples of SED fits to the NIRCam + ALMA SEDs. The observed photometry are shown as black dots and the $1\, \sigma$ errors are shown, including for the ALMA non-detections. Three types of models are fit: a) galaxy-only, without using the ALMA constraints (light blue), b) galaxy-only, with ALMA (blue), 3) AGN-only with hot dust (red), including ALMA. The left panel shows the \emph{HST+JWST} SED, while the middle panel shows the same fits including the IR-to-mm. The galaxy-only model is composed of three elements: 1) dust-free star forming component, 2) obscured star forming component, and 3) quiescent component, each with independent ages. Without ALMA the flexible stellar population model fits can reproduce the \emph{JWST} SED well, preferring highly reddened star formation. These fits overshoot the ALMA measurements by orders of magnitude in these cases. When including ALMA, the stellar only models lead to poor fits. ALMA limits the amount of dusty star formation, leaving only the old stellar population to reproduce the NIRCam colors. The AGN-only fits also require large amounts of dust, but the hot dust component of the AGN is fully consistent with the faint ALMA fluxes. The middle source is the brightest of the triply-lensed source presented in \citep{Furtak:2022}. Note the extremely blue UV slope of 23778 ($\beta \sim -4.0$), bluer than expected for normal stellar population models or AGN. 
}
\label{fig:galSED}
\end{figure*}

When we exclude the ALMA photometry, we are generally able to fit the SEDs with either composite galaxy or composite AGN models (Figure \ref{fig:galSED} left, Table \ref{table:thefits}). Figure \ref{fig:galSED} (left) illustrates that the galaxy-only fits without ALMA constraints nearly always prefer to fit the red end of the spectrum with dusty star formation, which can have a very similar slope to a reddened AGN continuum. The difference in $\chi^2$ between the AGN-only and galaxy-only fits are not significant in general. \citet{Barro:2023} also could not clearly distinguish between dusty star-formation and reddened AGN activity from their comprehensive spectral synthesis modeling to the compact red sources in CEERS. However, as we will show next, the predicted ALMA~1.2~mm fluxes for our adopted dusty star forming models are ruled out by the ALMA data, for typical star-forming galaxy dust templates. 

\subsection{ALMA Constraints}

We now perform SED fits including the ALMA photometry. The majority of the objects in the PSF-selected sample (20/\silvernr) have constraints from ALMA (see Table \ref{table:thefits}). The 1.2~mm measurement falls fairly close to the peak dust temperature for a dusty starburst at $z\sim5$. Even with substantial uncertainty on the dust peak temperature due to star formation, the ALMA point is constraining. Fitting to NIRCam only, the best-fit galaxy-only SED fits prefer a dusty star-forming component which over-predicts the ALMA 1.2~mm $2\sigma$ upper limits by a factor of a few to $>100$. Inclusion of the ALMA data leads to a significantly worse fit: ALMA limits the amount of dusty star formation and a quiescent stellar population struggles to reproduce the red NIRCam colors. The $\chi^2$ for the galaxy fit including the ALMA data is significantly worse than the galaxy-only fit excluding ALMA in all SED-selected objects, ranging from $\Delta \chi^2 = 6-340$. Dust-reddened AGN can also produce red NIRCam rest-frame optical colors, but are characterized by much hotter dust, peaking at an observed wavelength of $100 \micron$ \citep[rest wavelength $20~\micron$, e.g.,][]{Kirkpatrick:2012} and at least two orders of magnitude fainter at ALMA 1.2~mm than a dusty starburst with the same rest-optical spectrum (Figure \ref{fig:galSED}). 

\begin{figure*}
\includegraphics[width=0.90\textwidth]{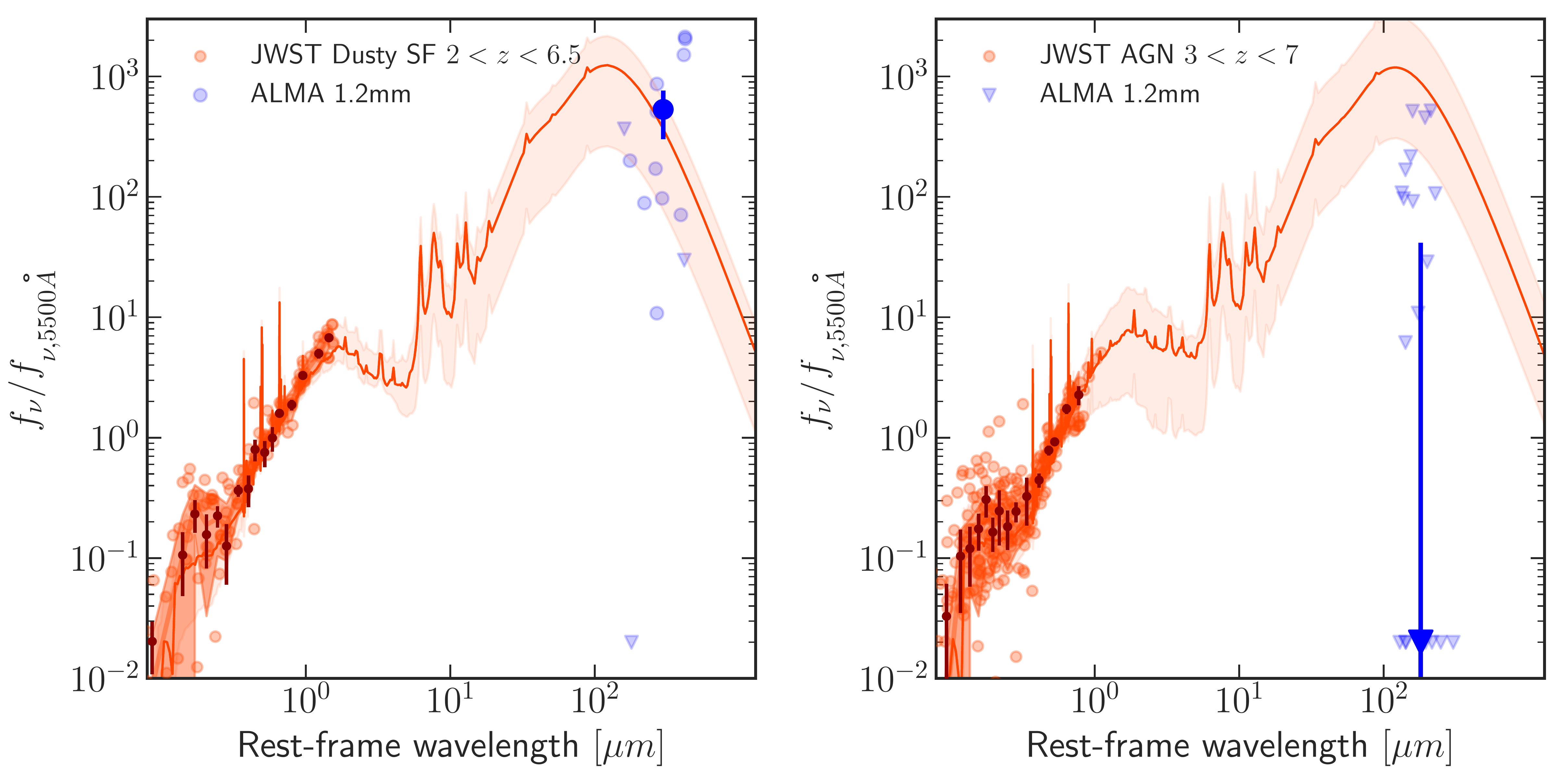}
\caption{Composite rest-frame SED of the 20/26 candidate AGNs with ALMA coverage (right) versus extended sources selected with the same color criterion but without requiring compactness (left). The extended sources are dusty star-forming galaxies at comparable redshifts. Red points indicate the individual HST+JWST photometry, shifted to the rest-frame using the photometric redshift. The red model shows the best-fit FSPS-based templates from \texttt{Eazy} where the IR SED is the predicted cool dust emission from star formation using the \citet{Draine:2007} dust emission model. The blue points are ALMA 1.2mm measurements with $2\sigma$ upper limits indicated as downward triangles. The large blue points are the stacked ALMA fluxes. Vertical lines denote $1\sigma$ uncertainties. The dusty SF sample has ALMA fluxes consistent with expectations for dust emission from star formation. In contrast, the compact candidate AGN sample has similar NIRCam colors, but is undetected in ALMA, with average fluxes $\gtrsim30$ times lower than expected for dust-obscured star formation. 
} 
\label{fig:almastack}
\end{figure*}

In Figure \ref{fig:almastack} right, we show the individual and stacked SEDs for the PSF-selected sample. The UV/optical SEDs are well-described by the dusty star-forming template from the template fits from \texttt{Eazy}. The ALMA points (no detections, shown are $2\sigma$ upper limits) fall below the predicted ALMA flux by a factor of a few to $>10$ per object, with the stacked $84\%$ upper flux limit $>30\times$ lower. In contrast, we also consider the ALMA measurements of the complementary star-forming sample from \S \ref{sec:DSFG}. These galaxies obey the same color cuts as the compact red sources but the sizes are extended rather than compact. We see that the NIRCam SEDs for both sources are very similar (by selection). However, the dusty star-forming galaxies are detected in ALMA in nearly all cases. Thus, the ALMA non-detections suggest that the compact red sources are not typical dusty star-forming galaxies, despite the galaxy-only fits being well fit with dusty star-formation in the rest-frame UV-optical.

We should mention the following caveats to our current dust templates. We fit with a fixed dust temperature distribution, corresponding to the default settings of FSPS $T_d \approx 20$~K \citep[e.g.,][]{Bethermin:2015,Jin:2019}. Compact star-forming galaxies at high redshift may have hotter dust temperatures \citep[e.g.,][]{Bethermin:2020, sommovigo2022, mauerhofer2023}. At these redshifts, the CMB temperature is also approaching these typical dust temperatures, and thus provides an additional source of heating \citep{daCunha:2015,Jin:2019}. As a test we reran the analysis with a dust temperature of $T_d\sim40$K, finding no significant changes in the results (see the appendix for more details). Unfortunately, we do not have enough long-wavelength information to perform more detailed dust modeling at present.

\subsection{Best-fit Models and the SED-selected Sample}
\label{sec:hiqual}

The SED modeling including ALMA upper limits prefers the AGN explanation because we do not detect the dust peak expected from cold dust in typical dusty star-forming galaxies. As a final test, we ask the fit to decide between AGN and galaxy templates, including star-forming (reddenend and unreddened), quiescent galaxies, and reddened and unreddened AGN templates. While the components are degenerate, these fits allow us to investigate in which cases the continuum {\it can} be fit with some combination of galaxy models without invoking any AGN. In general, the sources that have ALMA constraints cannot be fit without a dominant AGN component in the red. 

Based on these galaxy+AGN fits, we define the ``SED'' sample as the objects that have AGN-dominated fits at F444W. Specifically, in \goldnr\ sources, the AGN component comprises more than twice the flux of the galaxy component within the fit to the F444W band. In all but one case, the SED-selected targets are also the sources with ALMA constraints. 

From the reddened AGN component, we derive an implied bolometric luminosity from the dereddened 5100\AA\ luminosity, assuming that $L_{\rm bol} = 10 L_{5100}$ \citep[e.g.,][]{richardsetal2006,Shen:2020}. While imperfect, this model-dependent value represents our best guess for the intrinsic luminosity of the AGN in the context of our preferred AGN model. The inferred bolometric luminosities for the compact red sources range from $L_{\rm bol} \approx 10^{43}-10^{46}$~erg/s (Table \ref{table:thefits}).

\subsection{Noteworthy Objects}

There are a couple types of compact red sources worth calling out specifically.

\begin{figure}
\hspace{0.5cm}
\includegraphics[width=0.45\textwidth]{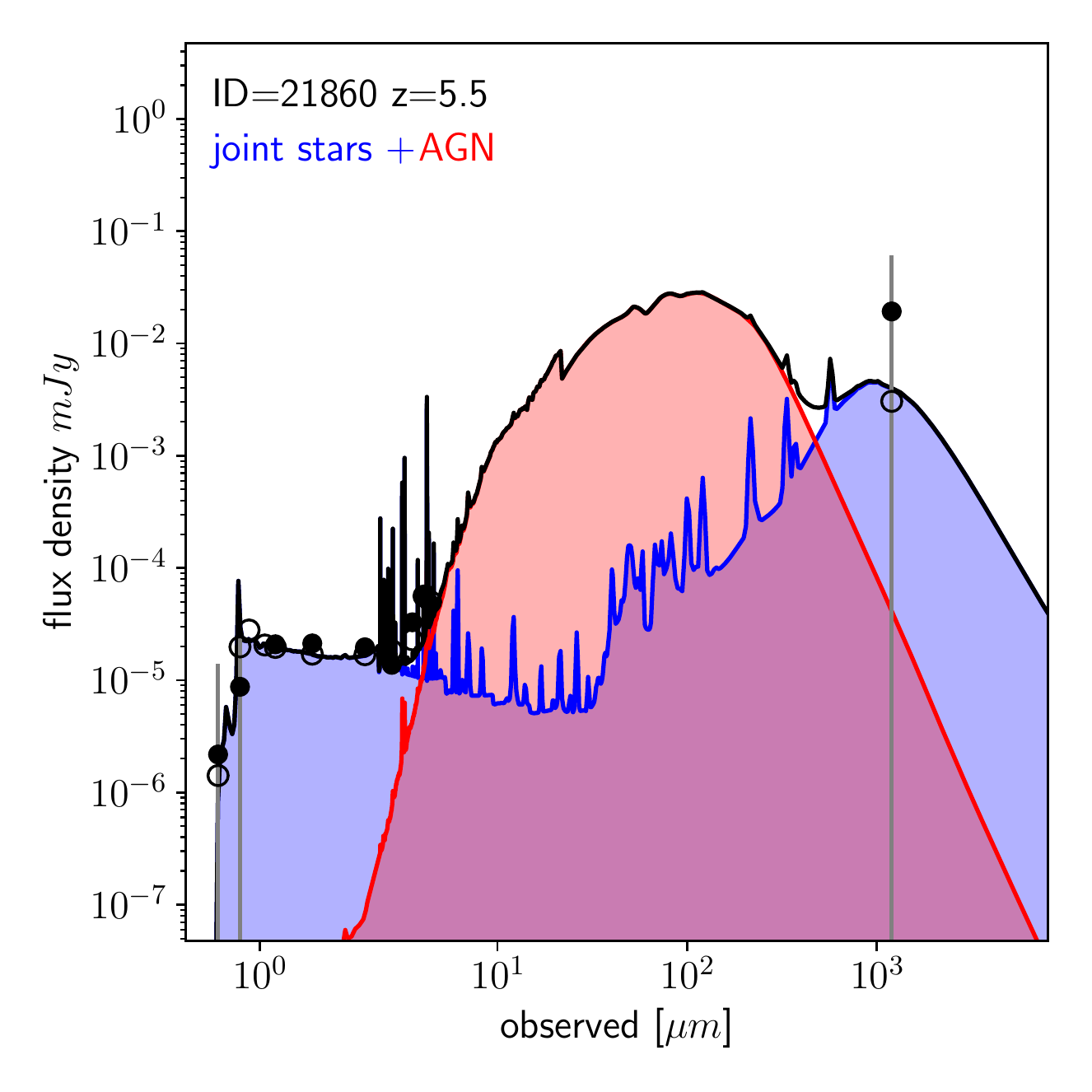}
\caption{A highly magnified compact red source. If the photometric redshift of $z \approx 5.5$ is confirmed, this source has an implied bolometric luminosity of $L_{\rm bol} < 10^{43}$~erg/s. This low luminosity could correspond to a very low black hole mass. As above, the photometric data are shown as black dots and the vertical lines are the $\pm 1\sigma$ uncertainties, the complete model is shown in black, and the model photometry is shown as open circles. The dominant sub-components for this fit are a blue galaxy component dominating the UV light and a heavily reddened AGN component dominating the rest-frame optical light. The ALMA limit severely limits the contribution from dusty star formation.}
\label{fig:lowL}
\end{figure}

Besides the Furtak object, we identify a strongly lensed source with $\mu = 40$ and a photometric redshift of $z=5.5$ (Figure \ref{fig:lowL}). We do not identify additional images for this object, as it lies just outside the multiply imaged region. Notably, however, the object lies in a highly sheared region. The tangential magnification for it is much higher than the radial one. Given that it appears to be a point source, this level of magnification could further tighten the constraints on its size once the redshift is secure. Also, if confirmed spectroscopically, this source would have an implied bolometric luminosity of $L_{\rm bol} < 10^{43}$~erg/s, making it one of the least-luminous AGN known at $z>4$. It would also be a strong candidate for a black hole with a mass \mbh$<10^6$~\msun. Black holes in this mass range are challenging to find locally, and only a handful of candidates exist at $z>0.3$ \citep[e.g.,][]{schrammetal2013,mezcua2017,mezcua2019,Halevi:2019,Greene:2020}. 

We also identify a handful of compact red sources that have extremely blue UV continuum slopes, with $\beta_{\rm UV} \approx -4$ where $f_{\lambda} \propto \lambda^{\beta}$. An example of this subset of objects is shown in Figure \ref{fig:galSED} (right; but see other sources with a green component in Figure \ref{fig:phot}). No quasar composite spectra have been observed to be this blue \citep[e.g.,][]{VandenBerk:2001,Davis:2007,Bonning:2007,Ivashchenko:2014,Temple:2021,Saccheo:2023}, with typical $\alpha_{\rm opt} = -0.5$ where $f_{\nu} \propto \nu^{\alpha_{\rm opt}}$, corresponding to $\beta_{\rm opt} = -1.5$ where $f_{\lambda} \propto \lambda^{\beta_{\rm opt}}$. These sources are thus simultaneously remarkably red and remarkably blue. Only spectra will be able to confirm our estimated photometric redshifts and (hopefully) determine the nature of these objects.

It is possible that the blue continuum reflects a scattered light component. Electron scattering would not change the color of the AGN spectrum, but dust scattering could, depending on the composition and geometry of the dust \citep[e.g.,][]{Kishimoto:2001,Draine:2003,Zakamska:2005}. Some indirect evidence exists for dust scattering in more luminous obscured AGN \citep{Kishimoto:2001,Alexandroff:2018}. In terms of starlight, in theory very young and low-metallicity stars may be able to get to $\beta_{\rm UV} = -4$ \citep[e.g.,][]{Schaerer:2003}, in practice the observed slopes of UV-selected galaxies have been closer to $\beta_{\rm UV} \approx -2$ \citep[e.g.,][]{Bouwens:2014} 


\subsection{Number Densities}


It is notoriously challenging to calculate effective volumes involving strong lensing \citep[e.g.,][]{Zitrin:2015,Bouwens:2017,Atek:2018,Ishigaki:2018}. Here, we adopt the lens model from \citet{Furtak:2022b_SLmodel} and the approach outlined in \citet{Atek:2018}. For each source, given the unlensed absolute magnitude, we calculate the effective volume of the survey by integrating over the area across the map with strong enough magnification to ensure we could detect that source. We do not account for varying incompleteness as a function of magnitude or photometric redshift, since the uncertainties about the intrinsic SEDs of these sources makes that modeling tricky. We do note that the sources are relatively bright ($>14 \sigma$) in F444W, which mitigates the magnitude incompleteness, and their compactness makes us less sensitive to surface brightness incompleteness. We do quantify systematic uncertainties in selection by comparing the number per bin for the Main sample of \silvernr\ objects and the SED-selected sample of \goldnr\ objects.

More specifically, our error bars include the following components: (1) Uncertainties from counting errors calculated using \citet{Gehrels:1986}; (2) Uncertainties from sample selection encapsulated by the difference in number between the main and SED-selected samples. When the SED-selected sample has no objects in a bin, we plot the $3 \sigma$ upper limit on zero sources as the upper error bar in that bin; (3) for the three bins with strongly lensed sources, we conservatively recalculate the volume in that bin with the lensed source removed.

We do not try to capture other systematic uncertainties associated with the lens modeling. In the Hubble Frontier Fields with the \emph{HST}-based lens models, there were multiple lens models, which afforded some kind of estimate of the impact of different decisions on lens model uncertainties. We do not yet have that suite of lens models based on the \emph{JWST} Abell 2744 data. We note that 22/\silvernr\ sources have fairly modest magnifications $\mu=1-4$, and lensing errors are not expected to impact the derived LFs. 

\begin{figure*}
\includegraphics[width=0.5\textwidth]{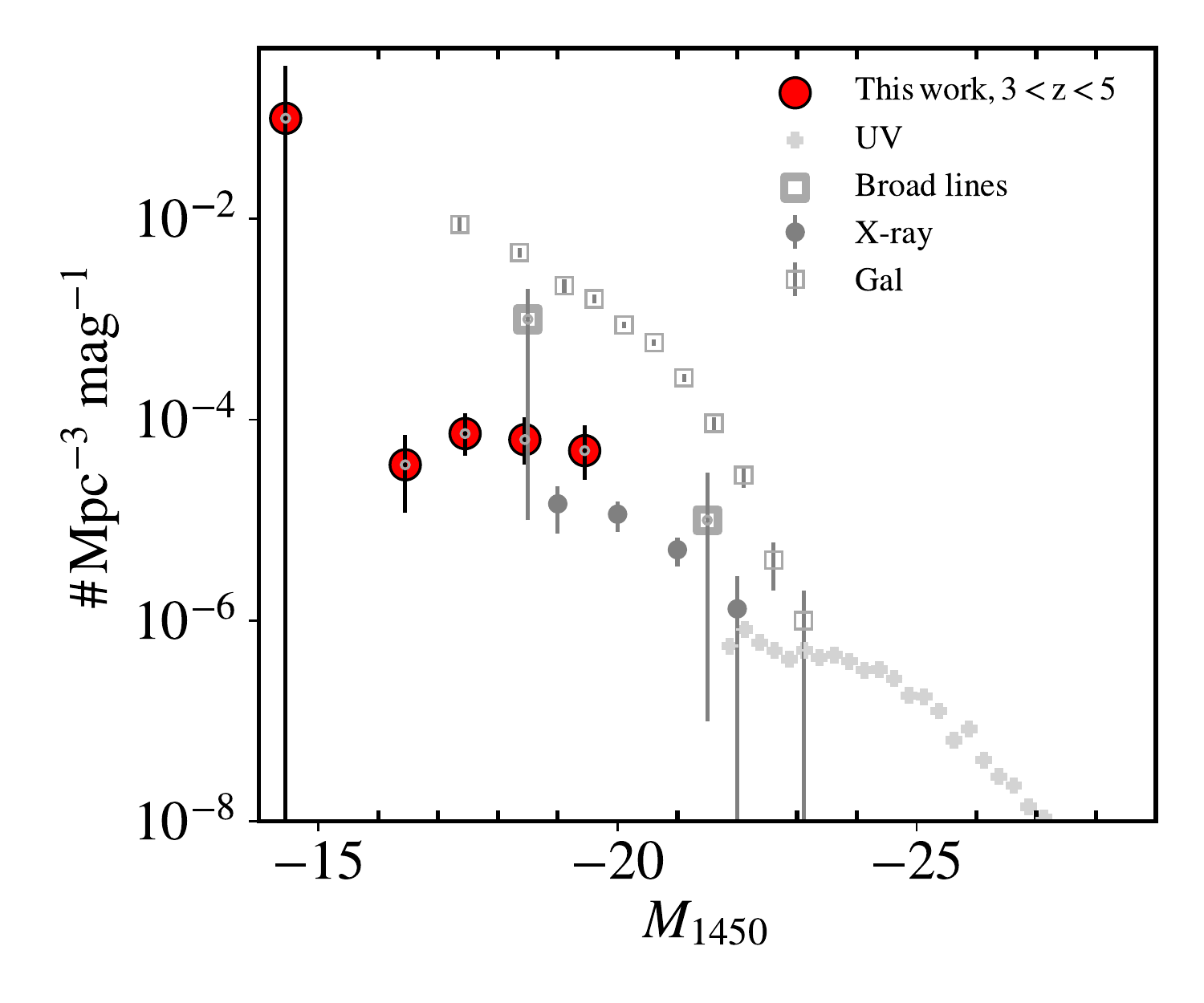}
\includegraphics[width=0.5\textwidth]{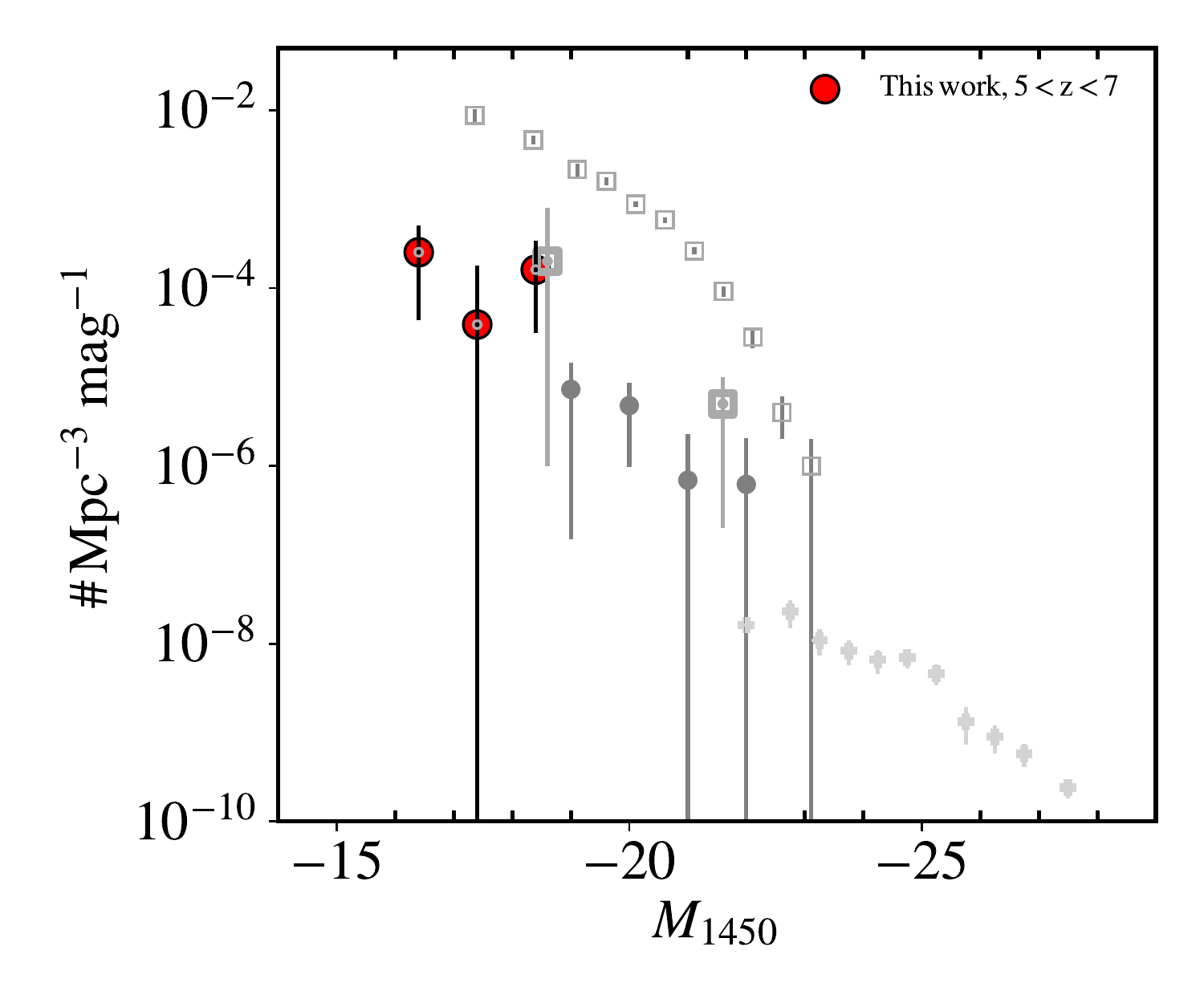}
\caption{{\it Left}: UV luminosity function for compact red sources with $3<z<5$, as compared with other samples from the literature. Compact red source luminosities have been corrected for magnification using the lensing model of \citet{Furtak:2022b_SLmodel}. The point below $M_{1450} = -15$ is from the single lensed source shown in Figure \ref{fig:lowL} and due to the strong lensing, the volume is quite uncertain. The UV AGN data are taken from \citet[][grey crosses]{Akiyama:2018}, the X-ray selected objects are presented in \citet[][dark grey circles]{Giallongo:2019}, and the galaxies are from \citet{Bouwens:2015}. \citet{Harikane:2022} presents the broad-line AGN from CEERS (thick grey boxes). We see that the number densities of the compact red sources is comparable to what is seen from X-ray selection, and broad-line selection as well.
{\it Right}: Same as left, but for lensing-corrected compact red sources between $5<z<7$. The UV AGN luminosity function is presented by \citet{Matsuoka:2018}; all other sources are as at left.
}
\label{fig:lfuv}
\end{figure*}

We present number densities in the rest-UV with $M_{1450}$, which is the standard for blue quasars (Figure \ref{fig:lfuv}). The UV luminosities are low compared to UV-selected samples, and we find much higher number densities than a naive extrapolation of the bright end from SDSS \citep[e.g.,][]{McGreer:2013,Finkelstein:2022,Fan:2023} and at lower luminosity from the Hyper-Suprime Camera sample \citep{Matsuoka:2018,Akiyama:2018,Matsuoka:2023}. Interestingly, \citet{Laporte:2017,Morishita:2020,Fujimoto:2022} also find hints that the number-density of UV-faint AGN could be much higher at low UV luminosity. Our number densities fall below the galaxy luminosity function \citep[here from ][]{Bouwens:2015}, and are consistent with the number-densities from broad-line AGN that have been spectroscopically identified in CEERS \citep{Harikane:2023,Barro:2023}.  If we deredden the UV luminosities, they would extend to $M_{1450} \approx -23$~mag. They would overlap with the faint-end of the AGN UV luminosity function \citep[e.g.,][]{Akiyama:2018,Matsuoka:2018,Matsuoka:2023}. At $\sim 10^{-5}$~Mpc$^{-3}$~mag$^{-1}$, the compact red sources presented here would be more than ten times more numerous than known UV-selected faint quasars.

It is hard to compare with the X-ray number counts, since they differ from survey to survey \citep[e.g.,][]{Aird:2015,Vito:2018,Giallongo:2019}. Our numbers are nominally consistent with, but on the high side of, these X-ray studies. It is worth saying that we may not be probing the same objects, since the existing X-ray counts require \emph{HST} cross-matches to enable photometric redshifts.

\section{Nature of Compact Red Sources}

We have presented a sample of \silvernr\ point-source dominated sources that are red in the rest-frame optical but have a UV component. From the \emph{HST}+NIRCam/\emph{JWST}+ALMA photometry alone, there are three primary interpretations of these sources. First, they could be red due to a strong Balmer/4000\AA\ break, and we could be seeing the cores of massive galaxies today \citep[e.g.,][]{Labbe:2023,Baggen:2023}. In some cases, the compact red sources have NIRCam LW colors that are too red to be fit with a 4000\AA\ break. Instead, galaxy-only templates tend to prefer a second solution with the red continuum arising from dusty star formation. These models tend to overpredict the ALMA upper limits by factors of 10 or more. The easiest way to have consistency with the data is to posit quite hot dust ($\gtrsim 100$K). Since the sources are very compact, it seems possible in principle that they have abnormally hot dust powered by star formation. The UV component could still be a weak UV-emitting AGN, and the red component dominated by dusty star formation \citep[e.g.,][]{Kocevski:2023}.

We have argued for a third possibility, based on the compact morphology and SED properties. We suggest that these sources are possibly dominated by reddened broad-line AGN with $3<z<7$. Rest-UV continuum could arise either from some star formation or through percent-level scattering of the unabsorbed AGN continuum. In the following two sections, we explore more quantitatively the implications should the compact red sources be dominated by dusty star formation or AGN emission respectively.


\subsection{Dusty Star formation Interpretation}

If the galaxies do not contain AGN but the observed SED is the result of dust-obscured star formation, the implications for the formation of massive galaxies could be significant. Few of these systems are bright enough to have been included in previous \emph{HST}-based surveys, or did not appear massive due to the shallow nature of the Spitzer/IRAC photometry \citep[e.g.,][]{Labbe:2023}. From our galaxy-only fits, several sources have inferred stellar masses of $> 10^{10}-10^{-11}$M$_\sun$ (Table \ref{table:thefits}). Their number densities are on the order of $\sim10^{-4}-10^{-5}$Mpc$^{-3}$ implying a substantial population that was overlooked at rest-UV/optical wavelengths before the advent of \emph{JWST}, echoing the results inferred from infrared/sub-mm studies \citep{Williams:2019,Casey:2021,Zavala:2021,Algera:2023, Fujimoto:2023} and \emph{HST}-dark galaxies \citep{Wang:2019,Barrufet:2023}.  Recent \emph{JWST} NIRCam/MIRI observations report the identification of point-source like candidate massive $> 10^{10}$M$_\sun$ galaxies at $z\sim8$ \citep{Akins:2023}, one of which may be detected at 2~mm, indicating the presence of a dusty starburst \citep{Akins:2023}. If these galaxies are all massive, extremely compact, and at high redshift, a straightforward interpretation would be that we are witnessing the formation of the dense compact inner regions of the most massive ellipticals today \citep{Baggen:2023}.

We note that the lack of a stacked ALMA 1.2 mm detection is unexpected if the reddest NIRCam wavelengths reflect typical dusty star-forming galaxies at these redshifts. Our complementary sample of dusty star-forming galaxies at $z=2-6$, selected with the same color selection as the compact red sources, have ALMA fluxes that are consistent with cool dust $T\sim20-40$K (see Figure \ref{fig:almastack}). Our sources are quite compact, and a relationship of increasing dust temperature with decreasing size has been found for SMGs \citep{Hodge:2016}, which would imply hotter $T_d\sim$60K for our sources. Dust temperatures based on Herschel+ALMA observations of $z\sim5-6$ drop-out selected galaxies are found to be warmer $T_d=40-60$K \citep{Bethermin:2020, sommovigo2022, mauerhofer2023}, and even higher $T_d\sim90$K have been suggested at $z\sim 7-8$ \citep{Laporte:2017,Behrens:2018,Bakx:2020}. Known examples of star forming galaxies with even hotter dust are rare. NGC1377 shows hot nuclear dust emission ($T_d > 180$K), but is likely powered by a buried AGN \citep{Aalto:2020}. Luminous IR sources with hot dust at high-redshift (hot-DOGS) also typically harbor powerful AGN \citep{Assef:2018}. Our observations do not preclude the existence of compact starbursts with high dust temperatures, although if they dominate our measured number densities then such systems are underrepresented in sub-mm/mm surveys of obscured star formation in the early universe \citep[e.g.,][]{Zavala:2021, Fujimoto:2023}.  


\subsection{AGN interpretation}



\begin{figure*}
\includegraphics[width=0.5\textwidth]{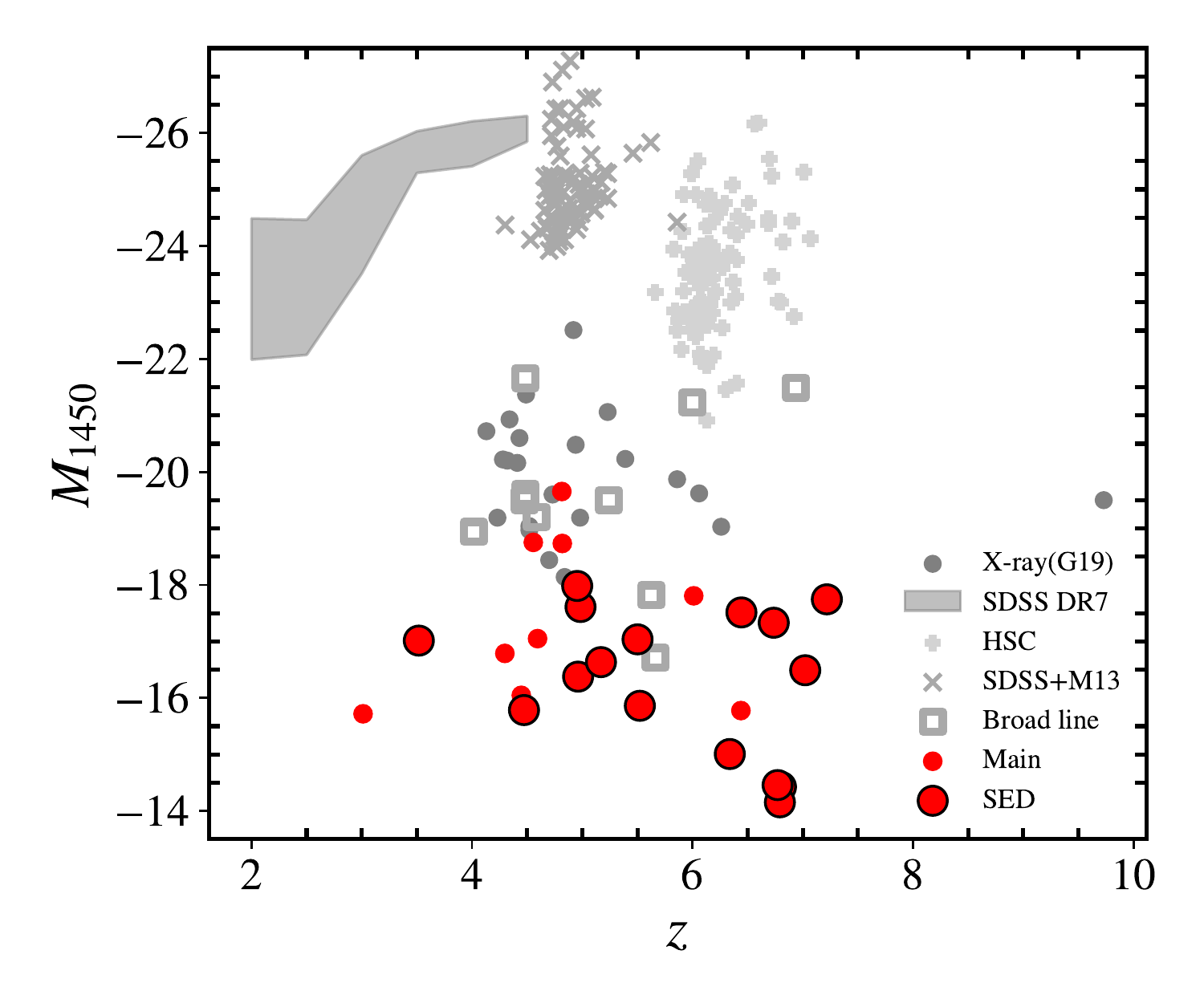}
\includegraphics[width=0.5\textwidth]{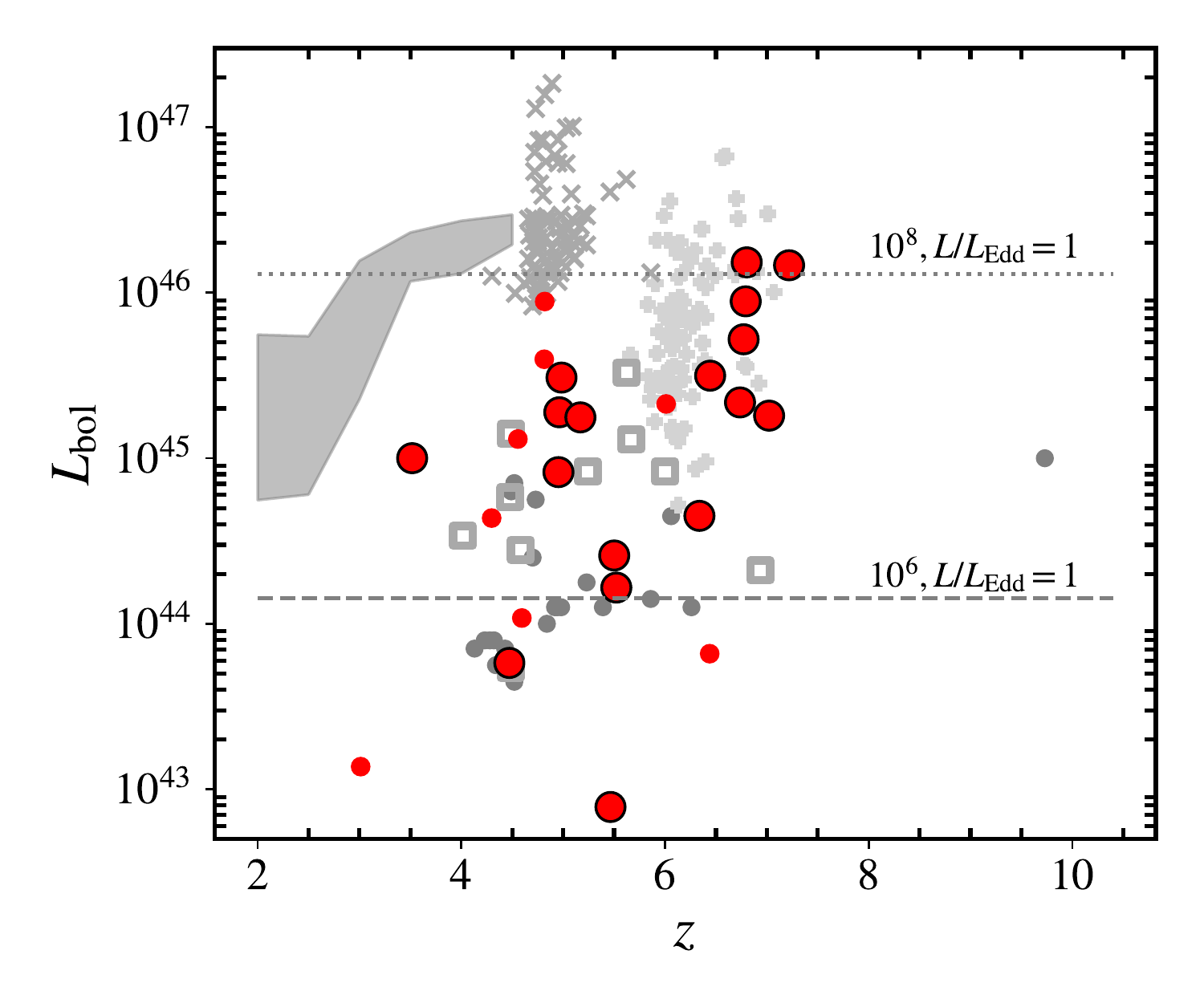}
\caption{{\it Left}: UV luminosities of the compact red sources (Main sample, red circles; SED-selected sub-sample identified with black dots) compared with various comparison samples from the literature, including SDSS DR7 \citep{Shen:2013}, and the high-redshift and luminous tail of SDSS quasars \citep{McGreer:2013}, the lower-luminosity quasars from the Hyper-Suprime Camera survey \citep{Akiyama:2018,Matsuoka:2018}, and the deep X-ray sample from GOODS-North \citep{Giallongo:2019}. The UV luminosities for the compact red sources are corrected for lensing using the model from \citet{Furtak:2023}, and only the high-quality sample have the black circles around them. The compact red sources lie at the very faint end of these samples when compared in the rest-frame UV at $\lambda=1450$\AA. 
{\it Right}: Bolometric luminosity distribution of the same comparison samples, showing that the compact red sources span the full range of observed quasar bolometric luminosities at these redshifts; luminosities also corrected for magnification using the \citet{Furtak:2023} model. We show the luminosity corresponding to a black hole at \mbh$=10^6,10^8$~\msun\ when accreting at the Eddington limit.
}
\label{fig:lz}
\end{figure*}

Pre-\emph{JWST}, it was common to characterize AGN at $z>5$ using the UV luminosity function, that being the only luminosity that we could measure. Our sources are very faint in the UV. Their magnification-corrected sources are fainter in the UV even than the lowest-luminosity AGN discovered via deep X-ray data \citep[Figure \ref{fig:lz} left][]{Giallongo:2019}. However, in our model the compact red sources are reddened, with $A_V \approx 2-6$. Thus, the UV luminosity does not represent the intrinsic luminosity of the accretion disk, as in unreddened AGN. Instead, the inferred bolometric luminosities derived from the dereddened rest-frame optical flux may be more illustrative of the true AGN luminosity.



What we find is that our targets span a wide range of bolometric luminosity, ranging from the Eddington limit for $\sim 10^6$~\msun\ black holes to $\sim 10^8$~\msun\ black holes. While we cannot infer black hole masses from the photometry, if we assume standard Eddington ratio distributions for these sources, typically $\sim 10\%$ of the Eddington limit \citep[e.g.,][]{Aird:2018}, then the typical source has \mbh$\sim 10^7-10^9$~\msun, similar to what was inferred spectroscopically by \citet{Kocevski:2023} and \citet{Harikane:2023}. 

It it useful to compare our targets with known populations of reddened broad-line AGN at lower redshift \citep[e.g.,][]{Glikman:2011,Glikman:2012,Banerji:2015,Assef:2018,Hamann:2017}. \citet{Banerji:2015} find that the reddened broad-line objects at $z \approx 2.5$ have higher number density than UV-bright quasars at the very highest luminosities $L_{\rm bol} > 10^{47}$~erg/s, but then their number begins to flatten at lower $L$. In contrast we are finding that our sources could outnumber their unreddened counterparts by a factor of ten or more.


We can also compare with X-ray searches at high redshift. In the UNCOVER field specifically, the emission from the hot gas in the cluster also complicates the analysis in the X-ray \citep{Bogdan:2023}, and will require more specialized analysis. However, the Little Red Dots presented in Matthee et al.\ are quite X-ray faint; we would likely need to reach ten-times deeper in the X-ray to find this population.

Given the very stringent limits on size measured for these sources, it is likely that the ratio of black hole to galaxy mass will be quite high for these targets. However, we prefer to defer a discussion of this topic to the time when we spectroscopically confirm the nature of the sources.

\section{Discussion and Summary}

Thanks to its remarkable red sensitivity, depth, and spatial resolution, \emph{JWST} has unveiled an unexpected population of very red and compact sources with photometric redshifts $z>3$ \citep{Endsley:2022,Labbe:2023,Furtak:2022,Akins:2023}. In a handful of cases, these compact red sources have been spectroscopically revealed to contain moderate-luminosity heavily reddened AGN \citep{Kocevski:2023,Harikane:2023,Ubler:2023,Oesch:2023}. Thus motivated, we searched systematically for such sources within the \emph{JWST} Treasury Survey UNCOVER \citep{Bezanson:2022}.  

We found a sample of \silvernr \ sources with red colors (F277-F444$> 1$ or F200-F356$>1$). Their light profiles are PSF-dominated, leading to typical lensing-corrected size limits of $<100$~pc at the presumed redshift range of $3 < z_{\rm phot} < 8$, and a median size of $\sim50$pc. These sizes alone do not preclude that the presence of massive very compact dusty star-forming galaxies. However, fits to the broad-band SEDs including ALMA can rule out typical dusty star formation with cool/warm dust in more than half the cases \citep[][in prep.]{Fujimoto:2023}. 

On net, we argue that a natural explanation for the bulk of the compact red sources would be that they are powered by heavily reddened AGN. In that case, they have comparable number densities to X-ray selected samples from the deepest X-ray data \citep{Vito:2018,Giallongo:2019} and they have implied bolometric luminosities spanning $10^{44} < L_{\rm bol} < 10^{46}$~erg/s. 
We find heavily reddened broad-line AGN at all redshifts, from the local Mrk~231 \citep{Veilleux:2016} and Mrk~1239 \citep[e.g.,][]{Pan:2021} to reddened AGN at cosmic noon \citep{Glikman:2012,Banerji:2015,Hamann:2017,Noboriguchi:2019,Noboriguchi:2022}. However, at $z<3$, highly reddened broad-line AGN with a UV component are rare. For instance, \citet{Noboriguchi:2019} estimate that 1\% of the HotDOGs that they select from the Hyper-Suprime Camera Survey \citep{Akiyama:2018} have a UV excess. 


\subsection{Exciting prospects for selecting AGN with JWST}

High-redshift quasars ($z>5$) have historically been selected via their blue continuum and Ly$\alpha$ break \citep[e.g.,][]{Fan:2023}. The final SDSS quasar search over 11,000 deg$^2$ uncovered a sample of 52 AGN at $z>6$. These luminous quasars at high redshift are very rare. Lower luminosity sources have been found using the deepest existing X-ray data, where near-infrared data are also required to determine galaxy counterparts and photometric redshifts for the X-ray sources. These lower-luminosity sources typically have higher number densities, but require very deep \emph{Chandra} data \citep[e.g.,][]{Aird:2015,Vito:2018,Giallongo:2019} along with deep NIR data to determine photometric redshifts.  

Even just limiting attention to our high-quality sample, we find a substantial number of probable AGN with $z>5$ in the $\sim 45$ arcmin$^2$ field. Identifying these targets to requires the combination of moderately deep \emph{JWST} data ($2-4$ hr per filter) and \emph{HST} data. It is clear that if these sources are confirmed, then there is a new and efficient way to identify AGN over a wide range of intrinsic luminosity through \emph{JWST}/NIRCam imaging. 
 
\subsection{Are most high-z AGN UV and X-ray faint?}

Deep X-ray searches combined with \emph{JWST} redshifts are finding new and very exciting sources even out to $z \approx 10$ \citep{Bogdan:2023}. However, the number densities of UV and X-ray selected AGN are rather low compared to the number densities of galaxies with $z>5$. For instance, taking local relations between black hole mass and galaxy mass, and reasonable assumptions for Eddington ratio distributions, \citet{VolonteriReines:2016} predict a higher number density of AGN than have been found with UV or X-ray searches. \citet{VolonteriReines:2016} invoke a lower ratio of black hole to galaxy mass for the bulk of the population. On the other hand, luminous quasars at these epochs appear to have higher ratios of black hole mass to galaxy mass \citep[see review in ][]{Fan:2023}. 

It is possible that UV-luminous quasars are a highly biased population in which the black holes have outgrown their galaxies, while the bulk of the black hole population at these redshifts are under-massive \citep[e.g.,][]{Zhang:2023}. Another possibility is that a large fraction of the AGN at these epochs are too red and obscured to be detected in the rest-frame UV by standard techniques, and too X-ray weak to be discovered in the X-ray. So far, neither the lensed \citet{Furtak:2022} object, nor the \citet{Kocevski:2023} objects are X-ray detected. Many models predict that the fraction of accretion that is obscured will grow with redshift \citep[e.g.,][]{Ni:2020}. So far, the empirical verdict on evolution of the obscuration fraction with redshift and luminosity has been a bit murky \citep[e.g.,][]{Aird:2015,Georgakakis:2015,Vito:2018,Giallongo:2019}. If we are discovering relatively obscured, moderate luminosity AGNs at high number density, our picture of the evolution in obscuration with redshift also may evolve. 

Along similar lines, \citet{Harikane:2023} use rest-frame optical spectroscopy to find that moderate luminosity AGNs account for $\sim 10\%$ of the galaxy population at $z>5$. Of these, $20\%$ of the broad-line AGN are in very red sources. This fraction is hard to interpret at the moment given the complicated selection function for spectra. A recent spectroscopic search using JWST NIRCam grism data revealed a substantial number of broad-line AGN in highly obscured compact objects at $z=5-6$ (Matthee et al. 2023). Certainly, early \emph{JWST} results suggest a higher number density of actively accreting black holes than was appreciated before. 


UNCOVER is a spectroscopic experiment. While the pre-imaging are in hand, the NIRSpec component is scheduled for July 2023. We hope to obtain spectroscopic identifications some of these targets at that time, to determine their nature as revealed by low-resolution spectroscopy in the UNCOVER region. Additional NIRCam grism spectroscopy and medium-band imaging of the UNCOVER field is scheduled in cycle 2, shedding additional light on the nature of these objects. Anything we find will be interesting in terms of understanding the nature of these targets. 



\clearpage

\section*{Acknowledgments}

J.E.G. and A.D.G acknowledge support from NSF/AAG grant\# 1007094, and J.E.G. also acknowledges support from  NSF/AAG grant \# 1007052. L.J.F. and A.Z. acknowledge support by Grant No.~2020750 from the United States-Israel Binational Science Foundation (BSF) and Grant No.~2109066 from the United States National Science Foundation (NSF), and by the Ministry of Science \& Technology of Israel. The Cosmic Dawn Center is funded by the Danish National Research Foundation (DNRF) under grant \#140. This work has received funding from the Swiss State Secretariat for Education, Research and Innovation (SERI) under contract number MB22.00072, as well as from the Swiss National Science Foundation (SNSF) through project grant 200020\_207349. PD acknowledges support from the NWO grant 016.VIDI.189.162 (``ODIN") and from the European Commission's and University of Groningen's CO-FUND Rosalind Franklin program. RPN acknowledges funding from {\it JWST} programs GO-1933 and GO-2279. Support for this work was provided by NASA through the NASA Hubble Fellowship grant HST-HF2-51515.001-A awarded by the Space Telescope Science Institute, which is operated by the Association of Universities for Research in Astronomy, Incorporated, under NASA contract NAS5-26555. The research of CCW is supported by NOIRLab, which is managed by the Association of Universities for Research in Astronomy (AURA) under a cooperative agreement with the National Science Foundation.  


\appendix
\section{SED fitting}
\label{appendix:sed}
To perform SED fitting with multiple components, e.g., stellar populations with potentially different reddening or AGN contribution, custom fitting is needed. 

For the AGN we use an empirical base model based on composite optical spectra from 2200 SDSS quasar spectra from \citep{VandenBerk:2001} and 27 quasar spectra observed with IRTF in the near-infrared by \citep{Glikman:2006}. This template is then reddened by $A_V=$0.5 to 5 using a \citet{Calzetti:2001} attenuation law.  When our fits include the ALMA 1.2~mm point, we model the mid-to-far infrared emission with the CLUMPY torus models \citep{Nenkova:2008mod,Nenkova:2008obs}. These models are implemented within the Flexible Stellar Population Synthesis (FSPS) code \citep{Conroy:2009}, with the torus thickness, number of clumps, power-law density of the clump distribution held constant at default values. The optical depth of a given clump is fixed at $\tau_{\rm AGN}=10$. The orientation of the torus is also fixed at $40\deg$, which means that virtually no emission from the AGN accretion disk is transmitted, and there is no AGN emission in the optical from this component of the model. More thorough discussion can be found in \citet{Leja:2018}. We tie the UV/optical disk component to the mid-to-far infrared component by scaling the torus model by the luminosity of the obscured component, thus ensuring energy balance. In some cases we wish to model both the blue rest-UV and red rest-optical wavelengths of the sources by AGN, where the blue light could reflect scattered light. In that case we use two components, one dust-free and one reddened, each with a logarithmic prior in luminosity from $13 < L_{bol} < 48$. Additional freedom is allowed in the broad lines and narrow lines, where the equivalent widths are modified by a small factor with a Gaussian prior with a dispersion of $0.3$ dex.


For the stellar populations we use three independent components based on FSPS: a dust-free star forming component, a dust obscured star forming component, and an old quiescent component. The star forming components assume constant-SFH and include nebular emission lines. The quiescent component is modeled as an exponentially declining SFH with $\tau=10$ Myr. The stellar population model assumes a Chabrier (2003) IMF and 0.2$Z_\sun$ metallicity. The ages of all three components may vary independently, in logarithmic steps, from 10 Myr to a maximum age corresponding to a formation redshift of $z=12$. The attenuation of the dusty component assumes a \citet{Calzetti:2001} attenuation law, with $A_V=$0.5 to 5.5 and a uniform prior. The masses of each component are drawn from a uniform prior between $6 < log(M/M_\sun) < 12$
When modeling the ALMA data, we use dust emission models as implemented in FSPS \citep[see description in][]{Leja:2017}. These are based on \citet{Draine:2007} dust models, which are parameterized with three parameters. As we cannot constrain the shape of the dust emission spectrum, the default settings for FSPS are adopted. The minimum starlight intensity is fixed to $U_{\rm min} = 1.0$, the relative fraction of the dust exposed to the minimum starlight is fixed to $\gamma = 0.01$, and the PAH mass fraction $q_{\rm PAH} = 3.5$. Together these three parameters establish the shape of the dust emission spectrum. The default settings correspond to an effective dust temperature of T$\sim20$K. We also ran the analysis with an elevated effective dust temperature, to account for dusty star formation with intrinsically higher temperatures or to take into account the effect the elevated CMB (daCunha et al. 2013). While there is no simple mapping from the parametrization of dust emission in FSPS to a typical modified black body at a single temperature, we increase $U_{\rm min} = 20.0$ finding it approximately corresponds to a dust peak emission of T$\sim40$K. The results do not change significantly with a higher dust temperature.


The model fitting is done by constructing composite models, drawing priors for each parameter, and evaluating the Bayesian evidence using the nested sampling algorithm  \texttt{ nestle } in \texttt{Python}. From the models, predicted photometry is generated using \texttt{sedpy}, the log likelihood is calculated by comparing to the {\it HST}, {\it JWST}, and {\it ALMA} observations, and credible intervals on the parameters are calculated from the posterior samples. 



\section{Sample}
\begin{figure*}
\includegraphics[width=0.99\textwidth]{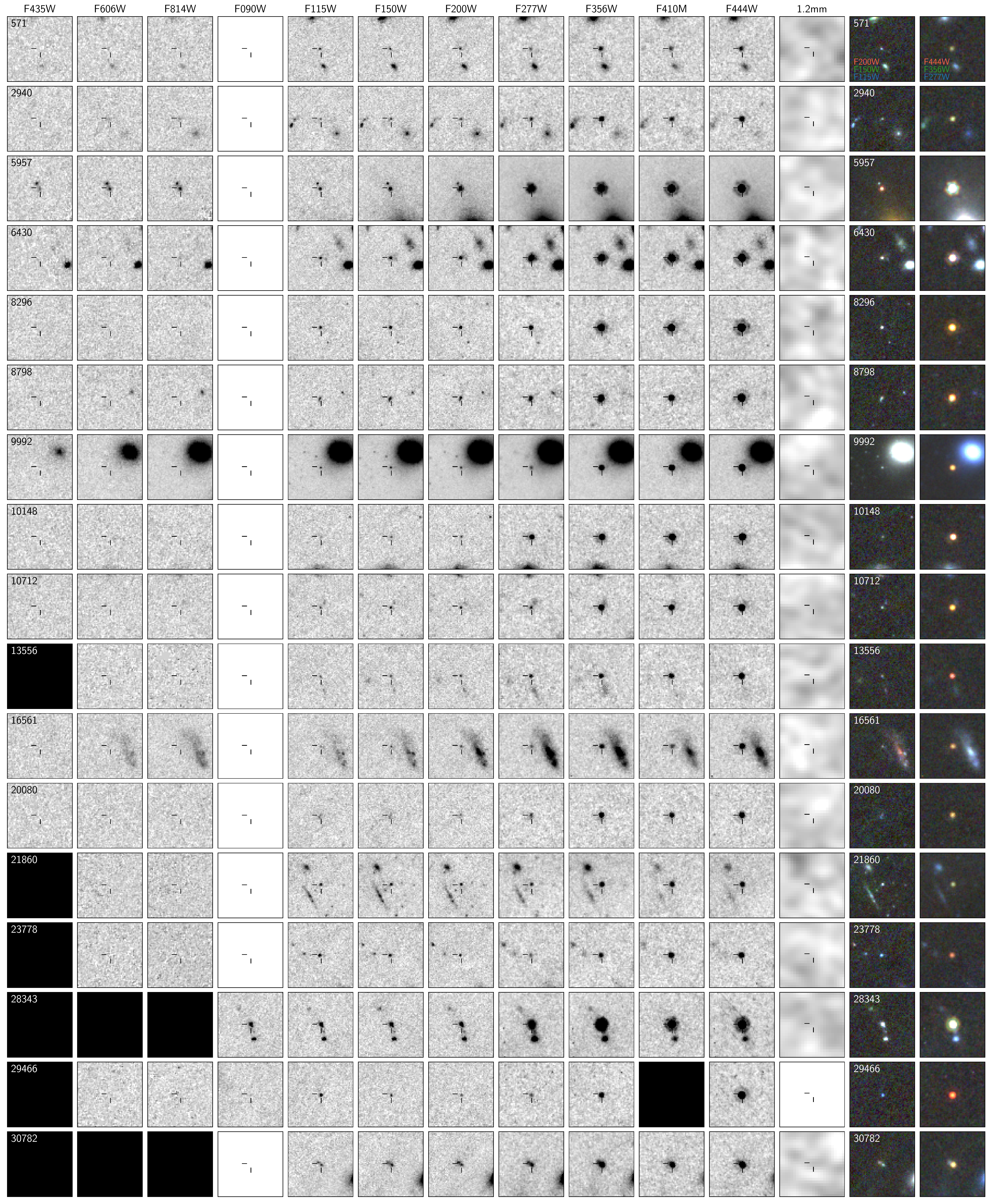}
\caption{Image stamps of high quality PSF + SED-selected sample}
\label{fig:appendix_stamps_gold}
\end{figure*}

\begin{figure*}
\includegraphics[width=0.99\textwidth]{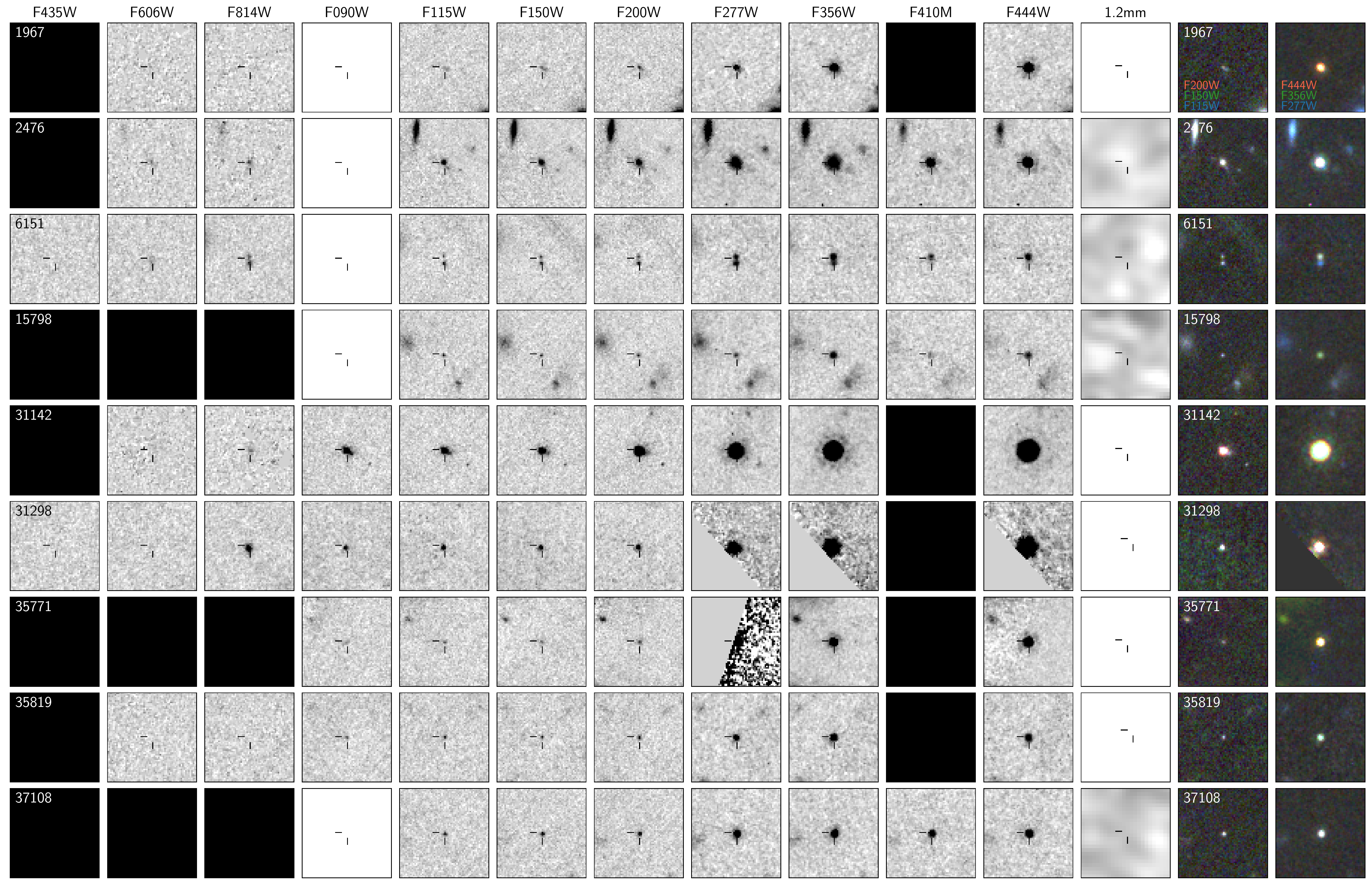}
\caption{Image stamps of the rest of the Main sample (PSF selected)}
\label{fig:appendix_stamps_silver}
\end{figure*}

\begin{figure*}
\includegraphics[width=0.7\textwidth]{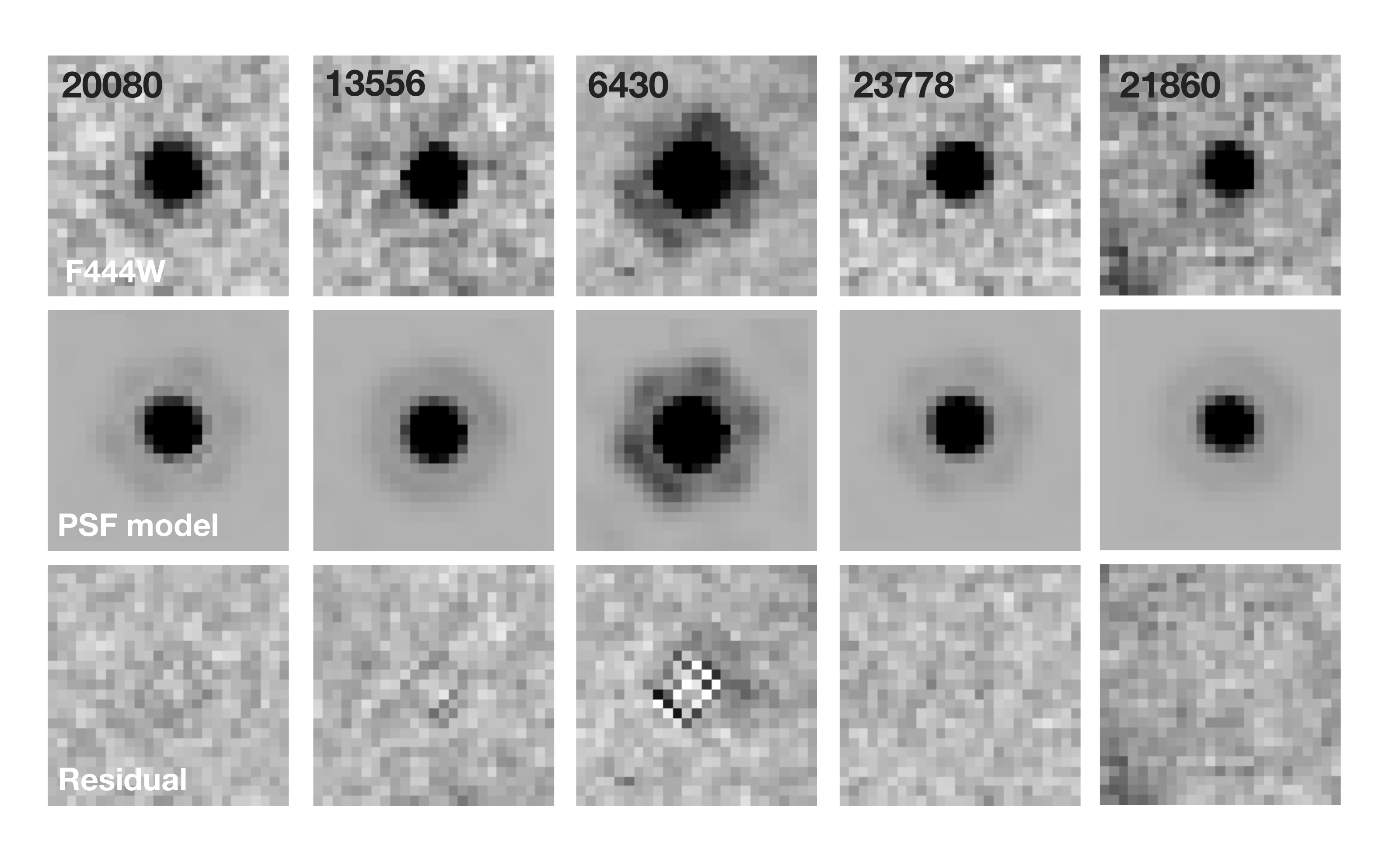}
\caption{Several examples of sources that are indistinguishable from a PSF. As a PSF model, a nearby star was selected and fitted and subtracted.}
\label{fig:appendix_psf}
\end{figure*}

\begin{figure*}
\includegraphics[width=0.99\textwidth]{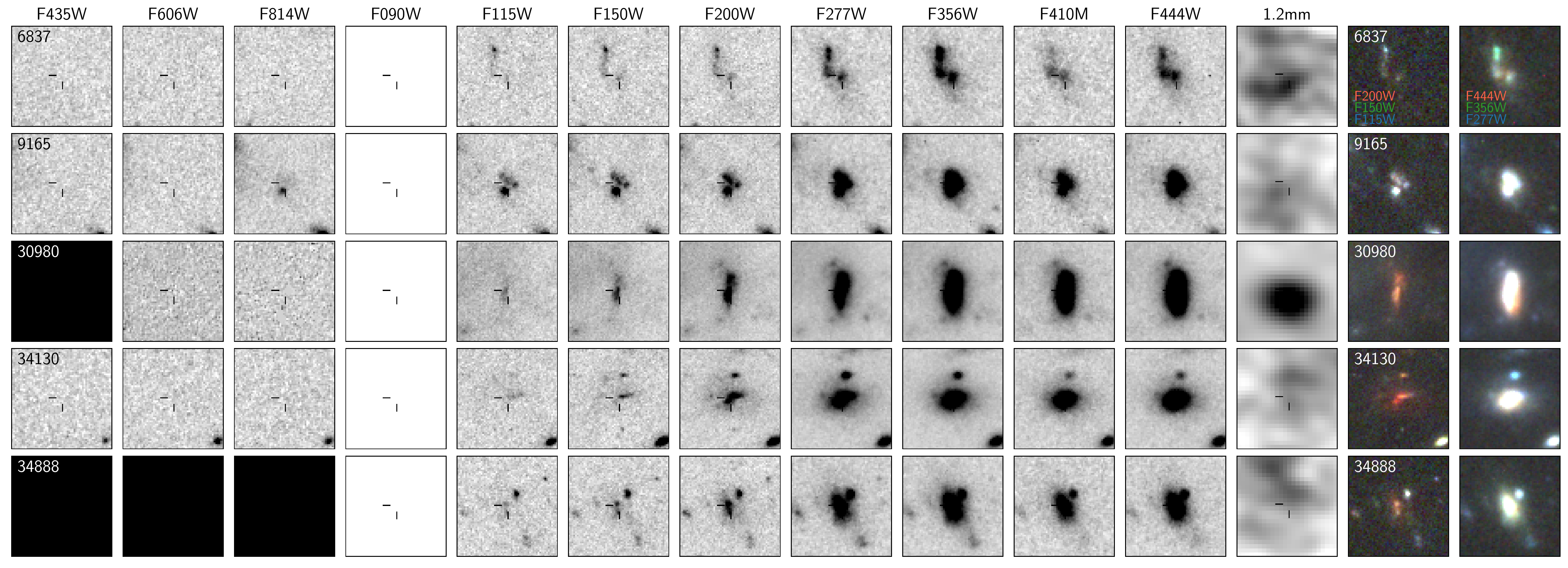}
\caption{Image stamps of examples of extended dusty galaxies}
\label{fig:appendix_stamps_dusty}
\end{figure*}

\begin{figure*}
\includegraphics[width=0.99\textwidth]{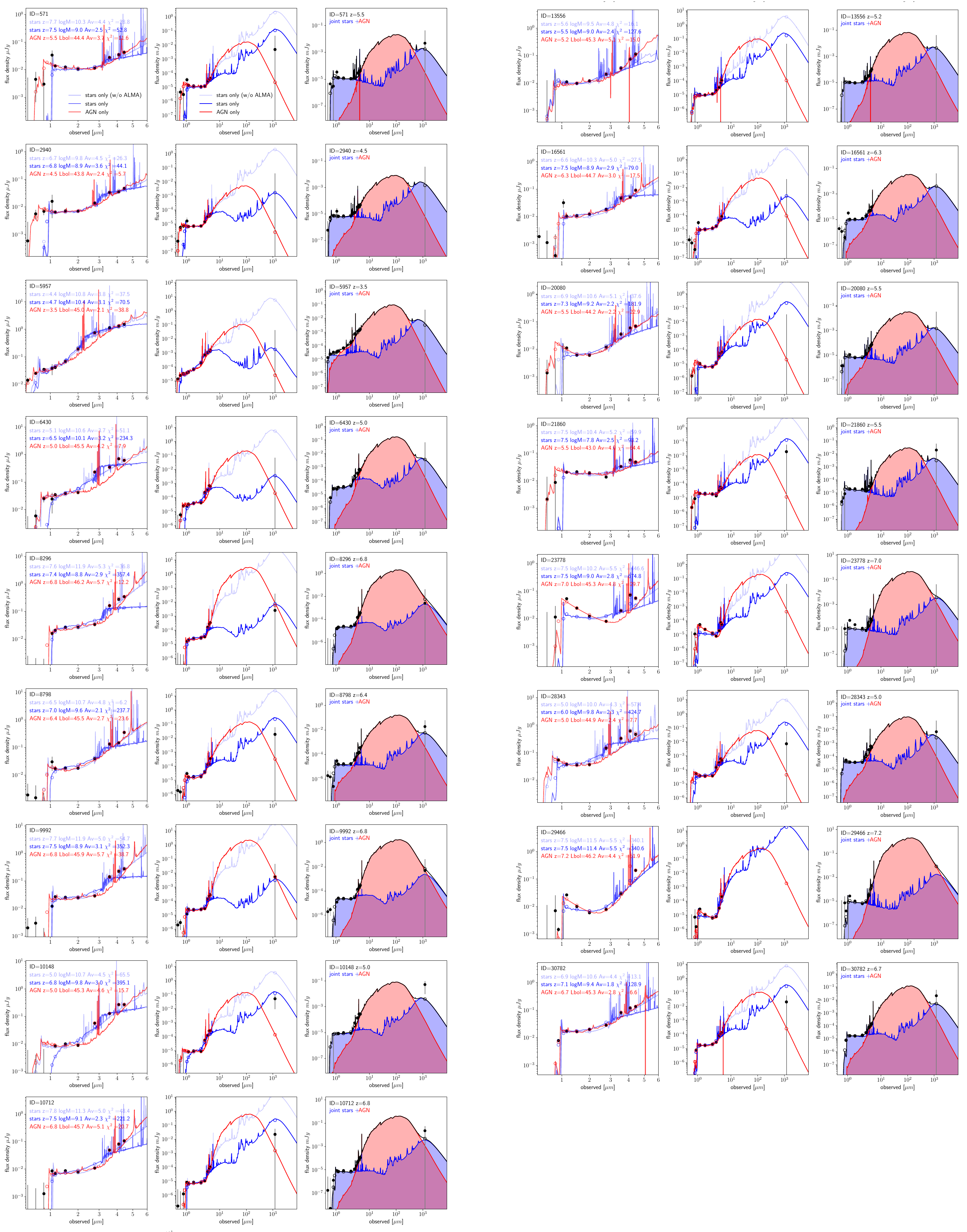}
\caption{PSF + SEDs fits of SED-selected sample}
\label{fig:appendix_sed_gold}
\end{figure*}

\begin{figure*}
\includegraphics[width=0.99\textwidth]{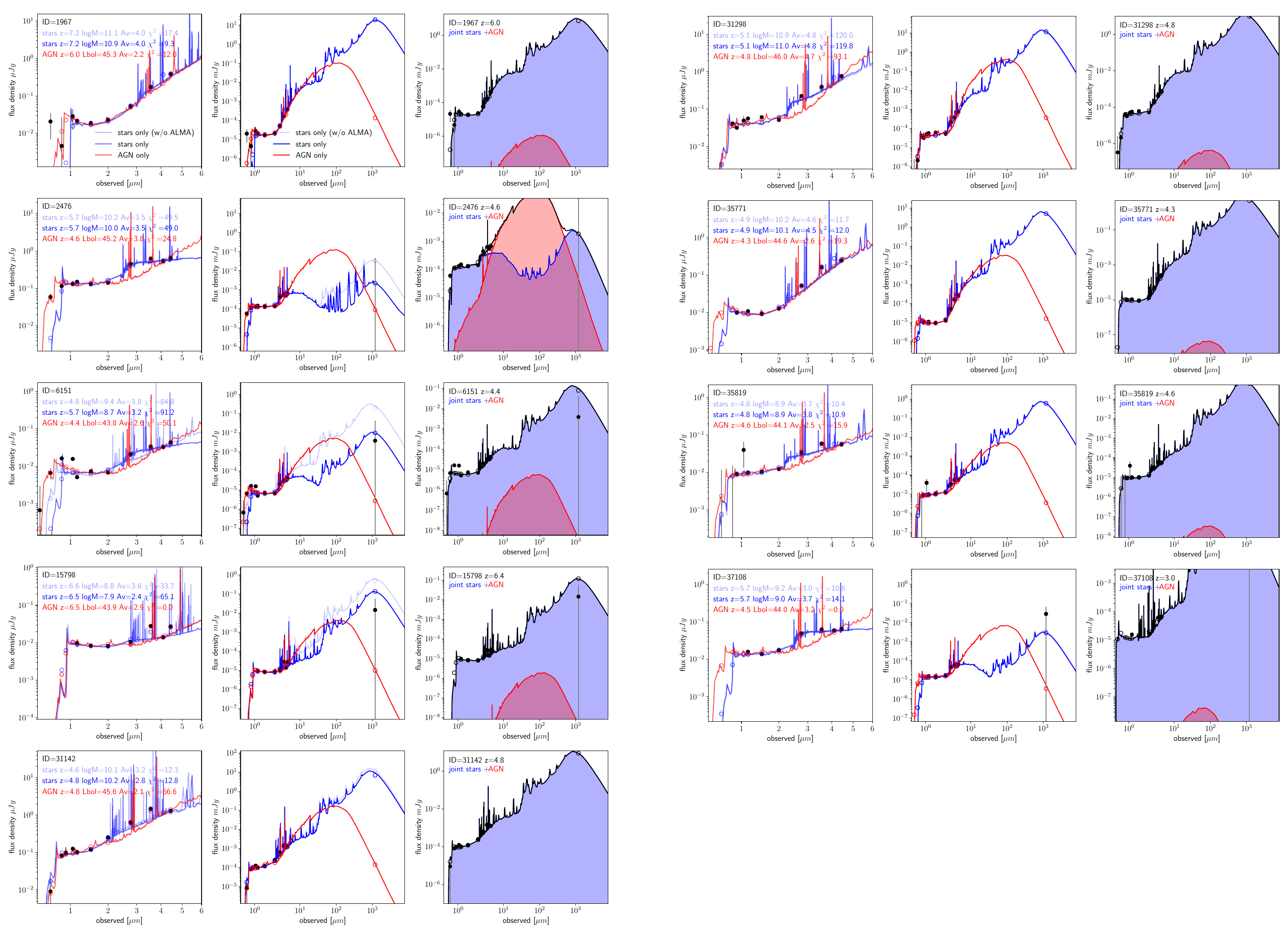}
\caption{SEDs of the rest of the Main sample}
\label{fig:appendix_sed_silver}
\end{figure*}



\end{document}